\documentclass[journal]{IEEEtran}
\usepackage{cite}
\usepackage{graphicx, epstopdf}
\usepackage{amsmath, amsfonts}
\usepackage{enumitem}
\begin{document}
\title{Single-Frequency Imaging and Material Characterization using Reconfigurable Reflectarrays}

\author{Weite Zhang,
        Hipolito Gomez-Sousa, Juan Heredia-Juesas,
        and Jose A. Martinez-Lorenzo, \it{IEEE Senior Member}
\thanks{W. Zhang is with the Department of Electrical and Computer Engineering, Northeastern University, Boston, MA, 02115 USA.}
\thanks{H. Gomez-Sousa, Juan Heredia-Juesas, and J. A. Martinez-Lorenzo are with the Department of Electrical and Computer Engineering and Department of Mechanical and Industrial Engineering, Northeastern University, Boston, MA, 02115 USA (e-mail: j.martinezlorenzo@neu.edu).}
}

\markboth{}
{}
\maketitle

\begin{abstract}
In this work, a physical and geometrical optics based single-frequency imaging scheme is proposed for personal screening systems using multiple reconfigurable reflectarrays. This scheme is able to not only reconstruct profiles of potential threat objects on human body, but also identify their materials in terms of their complex relative permittivities. Both simulation and experiment are carried out to detect dielectric objects at a microwave frequency of $\mathbf{24.16}$ GHz. The object profiles and complex relative permittivities are obtained with both high accuracy and computational efficiency, which show great potentials for security imaging where inspection of human body for threat materials, such as narcotics, explosives, and other types of contraband, is very common.
\end{abstract}


\IEEEpeerreviewmaketitle

\section{Introduction}
\IEEEPARstart{T}{he} use of electromagnetic (EM) waves in the microwave and millimeter-wave (mm-wave) bands has attracted intensive research interests during the past few decades in a variety of security\cite{Sheen2001Three, Ahmed2011A, Ahmed2012Advanced, Martinez2012SAR, Vaqueiro2014Compressed}, medical \cite{Meaney1996Microwave, Meaney2000clinical, Fear2002Confocal}, industrial \cite{Yujiri2003Passive, Benedetti2006Innovative, Sleasman2017Single}, and other important societal \cite{Bolomey1989Recent, Steinberg1991Microwave, Cloude1996Areview} applications. This is because its unique sensing and imaging capabilities. Specifically, at these frequency bands,  EM waves are non-ionizing--making them safe to be used in public spaces--and can be used to penetrate optically opaque materials, to create three-dimensional (3D) images, and to characterize and classify a wide range of hazardous materials, such as explosives related threats and other contraband or illicit substances and goods.

Conventional microwave and mm-wave radar imaging systems--such as those working in monostatic, bistatic, and multistatic configuraions\cite{Fear2013Microwave, Walterscheid2006Bistatic, Krieger2006Spaceborne, Herd2016The}--often require the use of a large bandwidth and stringent inter-antenna synchronization to enable coherent imaging, threat detection, and target classification. These characteristics not only  makes the imaging of frequency-dispersive objects more challenging, but also they substantially increase the complexity and cost of the multiple transmitting and receiving modules of the system. During the past decade, several imaging systems--based on compressive sensing (CS) theory \cite{Donoho2006Compressed, Alonso2010A, Vaqueiro2014Compressed}--have been proposed not only to reduce the hardware complexity but also to achieve a better imaging resolution when compared to that of traditional synthetic aperture radar (SAR) imaging systems\cite{Soumekh1999Synthetic}. However, CS algorithms still need intensive digital signal processing (DSP), setting a heavy computational cost at the receiving end, which ultimately precludes from their use in real-time imaging applications.

Recently, a reflectarray system has been able to perform real-time imaging in people-screening applications \cite{Paul2006Millimeter, Brendan2013Reflect, Gomez2017Modeling, abdillah2013identification}. The reflectarray is made of many 1-bit, phase-adaptable patch reflecting antennas \cite{Huang1991Microstrip, Pozar1997Design}, which enable multi-scale, beam focusing and imaging of targets located within a region of interest (RoI). Such a system annihilate the software-based computational cost of DSP imaging algorithms by the replacement with hardware-based focused imaging. Additionally, not only it produces high-resolution images but also operates in real-time; this is because the reflectarray is illuminated with a few transceiver antennas excited with a single-frequency continuous-wave (CW) signal. Notwithstanding, the single-frequency scheme only permits its use in near-field regions; which, in this case, has a maximum range of $\approx 2$ m.

Designing such reflectaray-based systems is challenging. This is due to the large electrical size of the reflectarray, so that predicting its performance capabilities in each focusing point of the RoI is computationally unfeasible. In \cite{Gomez2017Modeling}, an optimized physical optics (PO) method was proposed to simulate a single-reflectarray-based imaging system in a reasonable amount of time. The PO-based simulation platform was successfully used to reconstruct the profile of both dielectric and metallic objects. However, the object permittivity characterization and classification were not addressed in that work. Moreover, new societally-important emerging scenarios require to image even larger targets that, in some cases, may be distributed over a wide region\cite{baukus2000x, ahmed2011qpass, yurduseven2014indirect}; in these cases, additional reflectarrays are required to cover the entire imaging domain. Consequently, a more general PO-based method is needed to simulate multi-reflectarray screening systems.

Conventional EM security screening systems often posses high false alarm rates that ultimately result in uncomfortable pat-downs and reduced systems' throughput. One way to cut down these rates is by using the complex permittivities of objects to discriminate them between hazardous and innocuous materials. The complex permittivity can be characterized from the transmitted and received electromagnetic fields by different methods\cite{Gonzalez2013SAR, Alvarez2015SAR, Weatherall2016Spectral, Baginski2005Comparison, Zamani2017Estimation, Winters2006Estimation, Chan2015Material, Bourqui2016System, Mohammed2015Radar, Salman2012Determining, Islam2016Novel}. However, several drawbacks remain to be addressed before they can be efficiently used in realistic security applications. These include but are not limited to the following: (1) the need to use multiple transceivers or a large frequency bandwidth \cite{Gonzalez2013SAR, Alvarez2015SAR, Weatherall2016Spectral, Baginski2005Comparison, Zamani2017Estimation}, which may result in challenging detection and classification of frequency-dispersive objects \cite{Winters2006Estimation, Chan2015Material, Bourqui2016System}; and (2) the need to incorporate the object thickness \cite{Weir1974Automatic, Fenner2012A} or its borders\cite{Salman2012Determining, Islam2016Novel} as \textit{prior} information in the estimation process.

In this paper, a single-frequency imaging and material characterization method is proposed for multi-reflectarray systems, requiring no aforementioned object \textit{prior} information. This method is able to not only effectively and efficiently reconstruct the object profile, but also characterize the complex relative permittivity. During the permittivity estimation, which makes use of the range-dependent radiation pattern of the reflectarray in the near-field\cite{Nayeri2013Radiation} and considers both the magnitude and phase responses of the received fields, an accurate object thickness can be predicted by solving the phase-shift ambiguity\cite{Trabelsi2000Phase}. Such an ambiguity is unavoidable in conventional single-frequency characterization techniques\cite{Nayeri2013Radiation,Roelvink2012Measuring,You2017Free}, and can lead to a failure in discriminating any two Object-$i$, $i\in\{1,2\}$, that satisfy ${{T_1}\sqrt {{{\varepsilon_{1} '}}}  = {T_2}\sqrt {{{\varepsilon_{2} '}}} }$, $T_i$ and ${\varepsilon}_{i} '$ being the thickness and the dielectric constant of Object-$i$, respectively.

The rest of the paper is organized as follows: in Section \ref{sec_system_concept}, the concept of personal screening systems using multiple active reflectarrays is briefly described. In Section \ref{sec_imaging_theory}, a general imaging theory for reconstructing object profiles (using PO) and material characterization (using geometrical optics, GO) is derived. In Section \ref{section_results}, two-reflectarray-based computational simulations and experimental validations are carried out to detect dielectric slabs placed on the surface of a metallic plate at the a frequency of $24.16$ GHz. The results show the efficacy of the proposed method to image object profiles and estimate their complex relative permittivities. Section \ref{sec_conclusion} summarizes the conclusions of this work.

\section{System Concept} \label{sec_system_concept}
\begin{figure}[t]
\centering
\includegraphics[width=\linewidth]{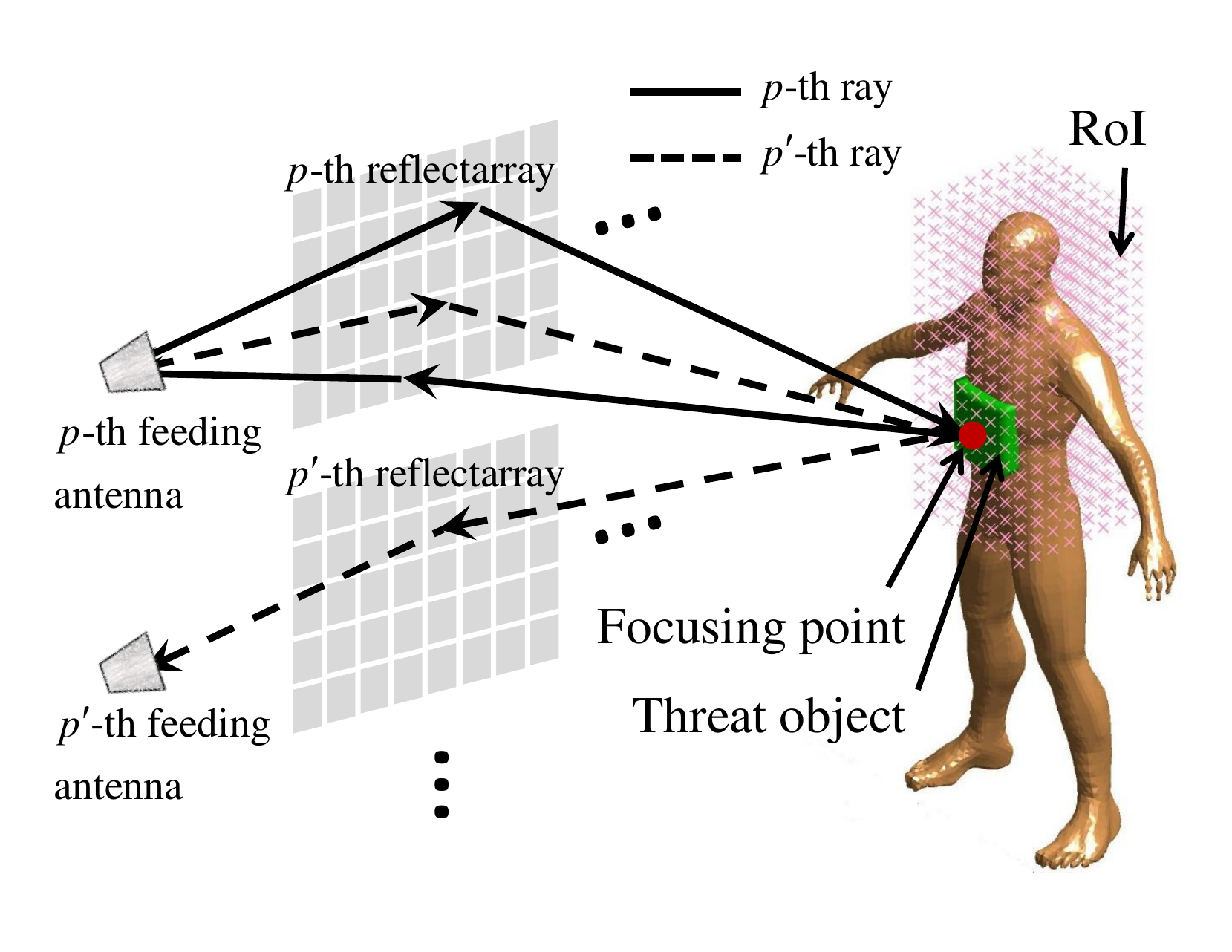}
\caption{System concept of microwave screening using multiple reflectarrays at $24.16$ GHz, where the $p$-th feeding antenna is active as a general analysis. The reflectarrays are all confocally configured to focus/refocus the single-frequency CW signal at a specific point, and let it scanned in the RoI. Each reflectarray has a corresponding feeding antenna for transmitting/receiving the radar signal. The concealed dielectric object under detection has an undetermined profile and complex relative permittivity.}
\label{system_model}
\end{figure}
The original idea of using multiple active reflectarrays in people-screening system was pioneered by Smiths Detection\cite{Paul2006Millimeter}. As is described in Fig. \ref{system_model}, the system has two principal components: (1) the feeding antennas that are used to transmit and receive the single-frequency CW radar signal; and (2) the reflectarrays that are used to focus the CW signal at a point in the RoI, when used in transmission mode, and to refocus the scattered field from that point into to the receiving antennas, when used in receiving mode. In this setup, a potentially concealed dielectric object in the RoI has an unknown profile and a complex relative permittivity of $\varepsilon_r = \varepsilon_r' - j\varepsilon_r''$, where ${{ \varepsilon }_r'}$ and ${{ \varepsilon }_r''}$ are the dielectric constant and loss factor, respectively.

Note that each reflectarray is equipped with a single horn antenna and transceiver module, and several reflectarrays can be confocally configured to simultaneously focus at the same point in space and to perform the imaging. In the general configuration shown in Fig. \ref{system_model}, all $P$ pairs of feeding-antenna and reflectarrays (FARAs) are used as follows. First, the CW signal from the $p$-th feeding antenna is used to illuminate its corresponding reflectarray. Then, this incident field is reflected from the reflectarray and focused at a point in the RoI; this is done by applying a binary phase to each reflectarray patch to make the free space propagation phase of each horn/patch/focusing-point ray to be as close as possible to zero \cite{Hum2005Realizing, Carrasco2012Reflectarray, Theissen2016binary}. Next, the focused incident field interacts with the object, thus producing a new field that is scattered towards the imaging system. Due to the confocal configuration, the scattered field is refocused through the $p'$-th receiving reflectarray towards its corresponding receiving antenna. Repeating the above procedure for all pairs of transmitting and receiving FARAs, an image of the target under test is finally created. It is important to note that the use of multiple FARAs provide a multiplexing gain that enhances the performance of the imaging system, when compared to the single FARA system described in Ref. \cite{Gomez2017Modeling}.

In the following Sections, the microwave operation frequency $f_0$ is selected to be 24.16 GHz. Because, at that frequency, clothing is essentially transparent, the human body is highly reflective, and dielectric materials are easily identifiable against the body \cite{Brendan2013Reflect}. The body is reasonably assumed to be a perfect electric conductor (PEC) plate.

\section{Imaging Theory} \label{sec_imaging_theory}
\subsection{Profile Reconstruction}
To get the simulated target profile, all surfaces of the feeding horn apertures, patches on the reflectarrays, and target are discretized into triangular facets. According to the exact free-space near-field equation described in \cite{Balanis2012Advanced}, the incident electric field ${\bf{E}}_{\mathrm{inc}}\left( {\bf{r}} \right)$ and magnetic field ${\bf{H}}_{\mathrm{inc}}\left( {\bf{r}} \right)$ at an observation point ${\bf{r}}$ can be computed using the electric ${{\bf{J}}}\left( {\bf{r'}} \right)$ and magnetic ${{\bf{M}}}\left( {\bf{r'}} \right)$ current sources, namely,
\begin{eqnarray}
\begin{aligned}
{{\bf{E}}_{{\rm{inc}}}}\left( {\bf{r}} \right) = & \int_S {\{  - {A_1}{G_1}{\bf{J}}\left( {{\bf{r'}}} \right) - {A_1}{G_2}\left[ {{\bf{J}}\left( {{\bf{r'}}} \right) \cdot {\bf{R}}} \right]{\bf{R}}} - \\
 &{B_1}{G_3}{\bf{M}}\left( {{\bf{r'}}} \right) \times {\bf{R}}\} {e^{ - j{k_0}R}}ds
\\
{{\bf{H}}_{{\rm{inc}}}}\left( {\bf{r}} \right) = &\int_S {\{  - {A_2}{G_1}{\bf{M}}\left( {{\bf{r'}}} \right) - {A_2}{G_2}\left[ {{\bf{M}}\left( {{\bf{r'}}} \right) \cdot {\bf{R}}} \right]{\bf{R}}} + \\
 &{B_2}{G_3}{\bf{J}}\left( {{\bf{r'}}} \right) \times {\bf{R}}\} {e^{ - j{k_0}R}}ds,
\end{aligned}
\label{eq_near_gen}
\end{eqnarray}
where $A_1 = \frac{{j}{\eta _0}}{{4\pi {k_0}}}$; $A_2 = \frac{{j}}{{4\pi {k_0} {\eta _0 ^2}}}$; $B_1 = B_2 = \frac{1}{{4{\pi}}}$; ${G_1} = \frac{{ - 1 - j{k_0}R + {{\left( {{k_0}R} \right)}^2}}}{{{R^3}}}$; ${G_2} = \frac{{3 + j3{k_0}R - {{\left( {{k_0}R} \right)}^2}}}{{{R^5}}}$; ${G_3} = \frac{{1 + j{k_0}R}}{{{R^3}}}$; ${\bf{R}} = {\bf{r}} - {\bf{r'}}$, $R = \left| {\bf{R}} \right|$; $k_0$ and $\eta_0$ are the wave number and impedance in free-space, respectively;  ${\bf{r}}$ is the observation point; ${\bf{r'}}$ is the source point; and $S$ is surface of the feeding antenna aperture.

With the incident fields ${\bf{E_{\mathrm{inc}}}}$ and ${\bf{H_{\mathrm{inc}}}}$, the induced electric ${\bf{J}}$ and magnetic ${\bf{M}}$ currents on any interface can be calculated using the modified equivalent current approximation (MECA) equations \cite{Meana2010MECA, gutierrez2011high}, which represents a generalization of the PO for both conducting and non-conducting dielectric surfaces:
\begin{eqnarray}
\begin{aligned}
{\bf{J}}\left( {\bf{r}} \right) = & \frac{1}{{{\eta_1}}}\left[ {E_{\mathrm{inc}}^{\mathrm{TE}}\cos {\theta _{\mathrm{inc}}}\left( {1 - {R_{\mathrm{TE}}}} \right){{{\bf{\hat e}}}_{\mathrm{TE}}}} \right. + \\
 &{\left. {\left. {E_{\mathrm{inc}}^{\mathrm{TM}}\left( {1 - {R_{\mathrm{TM}}}} \right)\left( {{\bf{\hat n}}_0 \times {{{\bf{\hat e}}}_{\mathrm{TM}}}} \right)} \right]} \right|_{{S_\mathrm{B}}}}\\
{\bf{M}}\left( {\bf{r}} \right) = & E_{\mathrm{inc}}^{TE}\left( {1 + {R_{\mathrm{TE}}}} \right)\left( {{{{\bf{\hat e}}}_{\mathrm{TM}}} \times {\bf{\hat n}}_0} \right) + \\
 &{\left. {E_{\mathrm{inc}}^{\mathrm{TM}}\cos {\theta _{\mathrm{inc}}}\left( {1 + {R_{\mathrm{TM}}}} \right){{{\bf{\hat e}}}_{\mathrm{TE}}}} \right|_{{S_\mathrm{B}}}},
\end{aligned}
\label{eq_boundary_condition2}
\end{eqnarray}
where the incident electric field is decomposed into its transverse electric (TE) and transverse magnetic (TM) modes, and analyzed individually; $S_B$ is the interface between the two media; ${\eta_1}$ is the wave impendence of the outwards medium; ${{{{\bf{\hat n}}}_0}}$ is the outward unit vector perpendicular to the interface; ${\theta_{\mathrm{inc}}}$ is the incident angle; $E_{\mathrm{inc}}^{\mathrm{TE}\mathrm{/}\mathrm{TM}}$ and ${{{\bf{\hat e}}}_{\mathrm{TE}/\mathrm{TM}}}$ are the incident field magnitude and the corresponding unit vector of the TE/TM mode, respectively; and $R_{\mathrm{TE}\mathrm{/}\mathrm{TM}}$ is the reflection coefficient of the TE/TM mode on the interface, which is defined as
\begin{eqnarray}
\begin{array}{l}
\begin{aligned}
{\left. {R_{\mathrm{TE}}} \right|_{\mu_1 = \mu_2}}  &= \frac{{\cos {\theta _{\mathrm{inc}}} - \sqrt {\frac{{{\varepsilon _2}}}{{{\varepsilon _1}}}}\sqrt {1 - \frac{{{\varepsilon _1}}}{{{\varepsilon _2}}}{{\sin }^2}{\theta _{\mathrm{inc}}}} }}{{\cos {\theta _{\mathrm{inc}}} + \sqrt{\frac{{{\varepsilon _2}}}{{{\varepsilon _1}}}}\sqrt {1 - \frac{{{\varepsilon _1}}}{{{\varepsilon _2}}}{{\sin }^2}{\theta _{\mathrm{inc}}}} }}\\
{\left. {R_{\mathrm{TM}}} \right|_{\mu_1 = \mu_2}} &= \frac{{ - \cos {\theta _{\mathrm{inc}}} + \sqrt{\frac{{{\varepsilon _1}}}{{{\varepsilon _2}}}}\sqrt {1 - \frac{{{\varepsilon _1}}}{{{\varepsilon _2}}}{{\sin }^2}{\theta _{\mathrm{inc}}}} }}{{\cos {\theta _{\mathrm{inc}}} + \sqrt{\frac{{{\varepsilon _1}}}{{{\varepsilon _2}}}}\sqrt {1 - \frac{{{\varepsilon _1}}}{{{\varepsilon _2}}}{{\sin }^2}{\theta _{\mathrm{inc}}}} }},
\end{aligned}
\end{array}
\label{eq_reflection_coe}
\end{eqnarray}
where ${\varepsilon_1}$ and ${\varepsilon_2}$ are the complex permittivity of the outwards and inner medium, respectively, and ${\mu_1}$ and ${\mu_2}$ are the corresponding permeabilities assumed to be equal. Note that on the surface of the PEC, magnetic current ${{\bf{M}}} = 0$ due to the fact that $R_{\mathrm{TE}}=R_{\mathrm{TM}}=-1$. Therefore, the magnetic currents on the feeding antenna apertures, the patch elements on the reflectarrays, and the human body are neglected.

\begin{figure}[t]
\centering
\includegraphics[width=\linewidth]{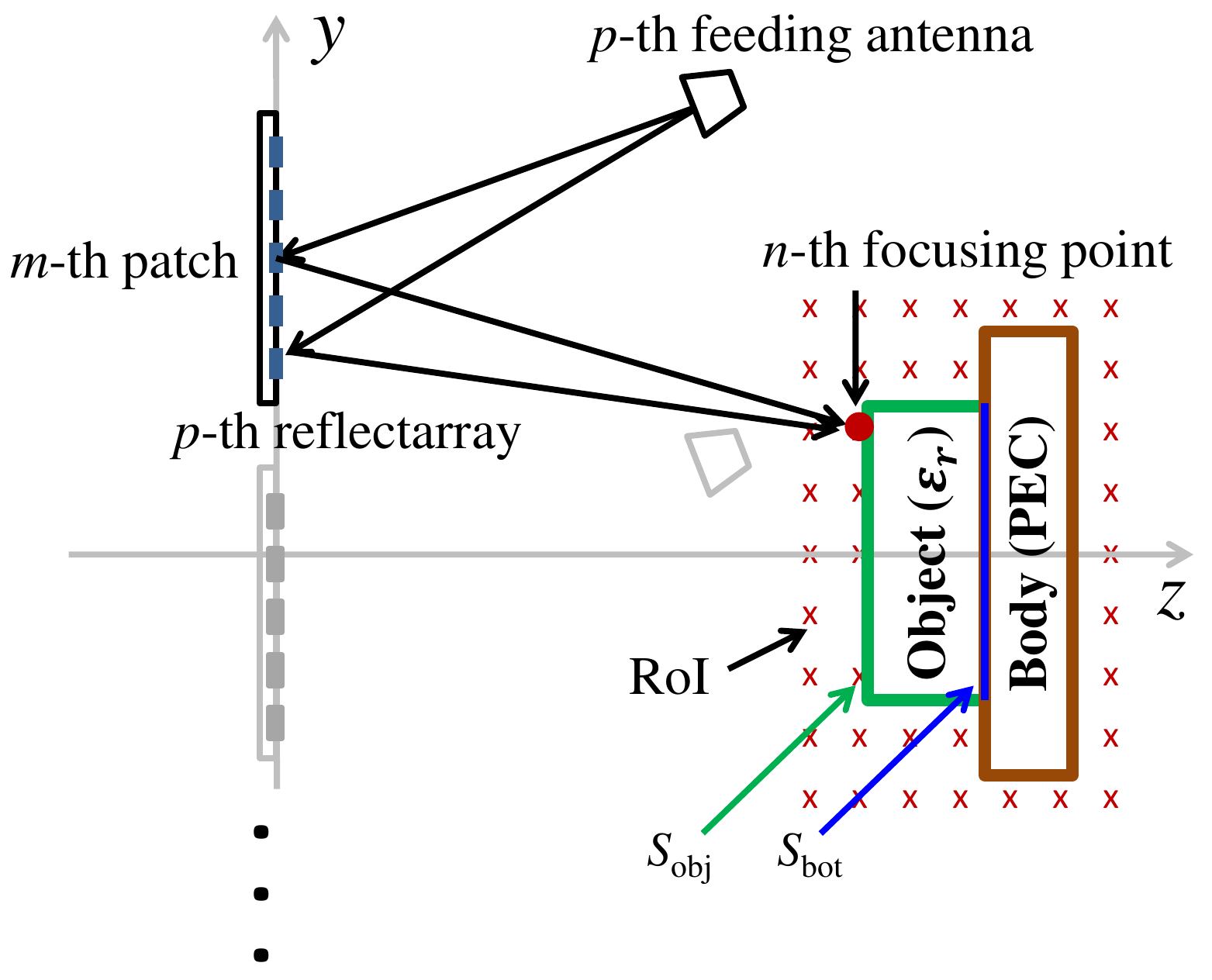}
\caption{Simulation model for the reconstruction of the threat profile. The body is assumed to be a PEC plate, and the threat, attached on the body surface, is a dielectric slab with undetermined profile and material in terms of the complex relative permittivity $\varepsilon_r$. $S_{\mathrm{obj}}$ and $S_{\mathrm{bot}}$ represent the air-object and object-body interface, respectively.}
\label{profile_reconstruction}
\end{figure}
As shown in Fig. \ref{profile_reconstruction}, assuming the total number of the patches on each reflectarray is $M$ and the electric current distribution of $p$-th feeding antenna is ${{\bf{J}}_{p}^{\mathrm{inc}}}$ $(p \in \{1,2,\dots,P\})$, the incident electric ${{\bf{E}}_{m,p}^{\mathrm{patch}}}$ and magnetic ${{\bf{H}}_{m,p}^{\mathrm{patch}}}$ fields on the $m$-th patch $(m \in \{1,2,\dots,M\})$ can be obtained using Eq. (\ref{eq_near_gen}). The corresponding induced electric current ${{\bf{J}}_{m,p}^{\mathrm{patch}}}$ can be calculated based on Eq. (\ref{eq_boundary_condition2}).

The reflectarrays are confocally set to focus the incident wave front to a desired point by using the binary phase approximation, namely, introducing a phase compensation {$\Delta \psi_{n,m,p}$} to each patch element,
\begin{eqnarray}
{\Delta \psi_{n,m,p}}  = \left\{ \begin{array}{l}
\pi , \quad \frac{\pi }{2} < {\rm{mod}}({k_0}\cdot{{L_{n,m,p}}},2\pi ) < \frac{{3\pi }}{2}\\
0, \quad {\mathop{\rm otherwise}\nolimits}
\end{array} \right.,
\label{eq_binary_phase_correction}
\end{eqnarray}
where ${\rm{mod}}(\cdot)$ is the modulus operator, and $L_{n,m,p} = \left( {\left| {{\bf{r}}_p^{{\rm{feed}}} - {{\bf{r}}_{m,p}^{{\rm{patch}}}}} \right| + \left| {{{\bf{r}}_{m,p}^{{\rm{patch}}}} - {\bf{r}}_n^{{\rm{focus}}}} \right|} \right)$. ${\bf{r}}_p^{{\rm{feed}}}$, ${\bf{r}}_{m,p}^{{\rm{patch}}}$, and ${\bf{r}}_n^{{\rm{focus}}}$ are the positions of the $p$-th feeding antenna, {the $m$-th patch of the $p$-th reflectarray}, and the $n$-th focusing point, respectively. Therefore, the electric current is modified as ${{\bf{J}}_{m,p}^{\mathrm{patch}}}{e^{j\Delta \psi_{n,m,p} }}$. Using Eq. (\ref{eq_near_gen}) again, we can calculate the incident electric ${{\bf{E}}_{m,n,p}^{\mathrm{target}}}$ and magnetic ${{\bf{H}}_{m,n,p}^{\mathrm{target}}}$ fields on the target surface. Thus, the total incident electric ${{\bf{E}}_{n,p}^{\mathrm{target}}}$ and magnetic ${{\bf{H}}_{n,p}^{\mathrm{target}}}$ fields for the $p$-th feeding antenna and the $n$-th focusing point are
\begin{eqnarray}
\begin{array}{l}
\begin{aligned}
{\bf{E}}_{n,p}^{{\rm{target}}} &= \sum\limits_{m = 1}^M {{\bf{E}}_{m,n,p}^{{\rm{target}}}} \\
{\bf{H}}_{n,p}^{{\rm{target}}} &= \sum\limits_{m = 1}^M {{\bf{H}}_{m,n,p}^{{\rm{target}}}}.
\end{aligned}
\end{array}
\end{eqnarray}
According to Eq. (\ref{eq_boundary_condition2}), the corresponding induced electric ${{\bf{J}}_{n,p}^{\mathrm{target}}}$ and magnetic ${{\bf{M}}_{n,p}^{\mathrm{target}}}$ currents can be written as
\begin{eqnarray}
\begin{array}{l}
\begin{aligned}
{\bf{J}}_{n,p}^{{\rm{target}}} &= {\bf{J}}_{n,p}^{{\rm{obj}}} + {\bf{J}}_{n,p}^{{\rm{body}}} \\
{\bf{M}}_{n,p}^{{\rm{target}}} &= {\bf{M}}_{n,p}^{{\rm{obj}}} + {\bf{M}}_{n,p}^{{\rm{body}}},
\end{aligned}
\end{array}
\label{eq_JM_target}
\end{eqnarray}
where ${{\bf{J}}_{n,p}^{\mathrm{\mathrm{obj}}}}$ and ${{\bf{M}}_{n,p}^{\mathrm{\mathrm{obj}}}}$ are the electric and magnetic currents, respectively, on the surface of the dielectric object. Similarly, ${{\bf{J}}_{n,p}^{\mathrm{body}}}$ and ${{\bf{M}}_{n,p}^{\mathrm{body}}}$ are the currents on the surface of the human body. Noticing that ${{\bf{M}}_{n,p}^{\mathrm{body}}} = 0$ for the human body (approximated to be PEC), one can calculate ${{\bf{J}}_{n,p}^{\mathrm{body}}}$ using $1^{\mathrm{st}}$-order PO method. While ${{\bf{J}}_{}^{\mathrm{\mathrm{}}}}$/${{\bf{M}}_{n,p}^{\mathrm{\mathrm{obj}}}}$ must be calculated considering multiple reflections within the dielectric object using a $K^{\mathrm{th}}$-order PO method based on Eq. (\ref{eq_near_gen}) and Eq. (\ref{eq_boundary_condition2}). The process can be described as
\begin{equation}
\begin{aligned}
&{\bf{J}}/{\bf{M}}_{n,p,0}^{\mathrm{obj}} \to {\bf{E}}/{\bf{H}}_{n,p,1}^{\mathrm{bot}} \to {\bf{J}}/{\bf{M}}_{n,p,1}^{\mathrm{bot}} \to {\bf{E}}/{\bf{H}}_{n,p,1}^{\mathrm{obj}}\\
 \to &{\bf{J}}/{\bf{M}}_{n,p,1}^{\mathrm{obj}} \to {\bf{E}}/{\bf{H}}_{n,p,2}^{\mathrm{bot}} \to {\bf{J}}/{\bf{M}}_{n,p,2}^{\mathrm{bot}} \to {\bf{E}}/{\bf{H}}_{n,p,2}^{\mathrm{obj}}\\
  \to &{\bf{J}}/{\bf{M}}_{n,p,2}^{\mathrm{obj}} \to {\bf{E}}/{\bf{H}}_{n,p,3}^{\mathrm{bot}} \to {\bf{J}}/{\bf{M}}_{n,p,3}^{\mathrm{bot}} \to {\bf{E}}/{\bf{H}}_{n,p,3}^{\mathrm{obj}}\\
 \vdots \\
 \to &{\bf{J}}/{\bf{M}}_{n,p,K-1}^{\mathrm{obj}},
\end{aligned}
\label{eq_JM_obj_multi}
\end{equation}
where ${\bf{J}}_{}^{\mathrm{}}$/${\bf{M}}_{n,p,0}^{\mathrm{obj}}$ is denoted as the electric or magnetic currents induced by the initial reflection on the air-object interface $S_{\mathrm{obj}}$; ${\bf{E}}_{}^{\mathrm{}}$/${\bf{H}}_{n,p,k}^{\mathrm{bot}}$ and ${\bf{J}}_{}^{\mathrm{}}$/${\bf{M}}_{n,p,k}^{\mathrm{bot}}$ are the incident fields and the induced currents, respectively, on the object-body interface $S_{\mathrm{bot}}$ after $k$ $(k \in \{1,2,\dots,K-1\})$ reflections within the dielectric object; Similarly, ${\bf{E}}_{}^{\mathrm{}}$/${\bf{H}}_{n,p,k}^{\mathrm{obj}}$ and ${\bf{J}}_{}^{\mathrm{}}$/${\bf{M}}_{n,p,k}^{\mathrm{obj}}$ correspond to the incident fields and the induced currents, respectively, on $S_{\mathrm{obj}}$. Consequently, the total electric ${\bf{J}}_{n, p}^{\mathrm{obj}}$ and magnetic ${\bf{M}}_{n, p}^{\mathrm{obj}}$ currents on $S_{\mathrm{obj}}$ are computed by
\begin{eqnarray}
\begin{array}{l}
\begin{aligned}
{\bf{J}}_{n,p}^{{\rm{obj}}} &= {\bf{J}}_{n,p,0}^{{\rm{obj}}} - \sum\limits_{k = 1}^{K - 1} {{\bf{J}}_{n,p,k}^{{\rm{obj}}}} \\
{\bf{M}}_{n,p}^{{\rm{obj}}} &= {\bf{M}}_{n,p,0}^{{\rm{obj}}} - \sum\limits_{k = 1}^{K - 1} {{\bf{M}}_{n,p,k}^{{\rm{obj}}}} .
\end{aligned}
\end{array}
\label{eq_JM_obj}
\end{eqnarray}

As all the reflectarrays are confocally arranged, all the receiving antennas are able to receive the scattered signal. Define ${\bf{E}}_{n,p}^\mathrm{rec}$ as the total received electric field from the $p$-th receiving antenna with all $P$ feeding antennas active and the beam focused at the $n$-th point. Although, ${\bf{E}}_{n,p}^\mathrm{rec}$ can be calculated via an inverse computational procedure using the PO method from the target to the receiving horns based on Eq. (\ref{eq_near_gen}) and Eq. (\ref{eq_boundary_condition2}), in order to improve the computational efficiency, the general reciprocity theorem for multiple-in-multiple-out (MIMO) systems is used as follows:
\begin{equation}
E_{n,p}^{{\rm{rec}}} = {\frac{{\int_S {{\bf{E}}_{n,p}^{{\rm{target}}}\cdot{\bf{J}}_n^{{\rm{target}}} - {\bf{H}}_{n,p}^{{\rm{target}}}\cdot{\bf{M}}_n^{{\rm{target}}}ds} }}{{\int_S {{\bf{\hat e}}_{n,p}^{{\rm{rec}}}\cdot{\bf{J}}_{p}^{{\rm{inc}}}ds} }}},
\label{eq_reciprocity_theorem}
\end{equation}
where $E_{n,p}^{{\rm{rec}}}$ is the amplitude of the received field $\mathbf{E}_{n,p}^{{\rm{rec}}}$, considered to be uniform on the receiving apertures; ${{\bf{J}}_{n}^{\mathrm{target}}} = \sum\limits_{p = 1}^P {{\bf{J}}_{n,p}^{\mathrm{target}}}$ and ${{\bf{M}}_{n}^{\mathrm{target}}} = \sum\limits_{p = 1}^P {{\bf{M}}_{n,p}^{\mathrm{target}}}$ are the total induced electric and magnetic currents on the target surface, respectively; and ${\bf{\hat e}}_{n,p}^{{\rm{rec}}}$ represents the uniform complex polarization vector of $\mathbf{E}_{n,p}^{{\rm{rec}}}$.

Finally, the target profile can be reconstructed by adding the received electric field $E_{n,p}^{{\rm{rec}}}$ from all $P$ receivers,
\begin{equation}
E_n^{{\rm{rec}}} = \sum\limits_{p = 1}^P {E_{n,p}^{{\rm{rec}}}},
\end{equation}
and determining the location of the maximum total received field along the $z$-axis, namely,
\begin{eqnarray}
z_n^{{\rm{imaging}}}\left( {{x_n},{y_n}} \right) = \mathop {\max }\limits_{{z_n}} \left\{ {\left| {E_n^{{\rm{rec}}}\left( {{x_n},{y_n},{z_n}} \right)} \right|} \right\},
\label{eq_profle_reconstruction}
\end{eqnarray}
where $\left( {x_n,y_n,z_n} \right)$ is the location of the $n$-th focusing point.
%denoting the imaged object center as $[{x_\mathrm{c}}, {y_\mathrm{c}}, z_\mathrm{c}]$, $z_\mathrm{c} = z_\mathrm{c}^{{\rm{imaging}}}\left( {{x_\mathrm{c}},{y_\mathrm{c}}} \right)$, which can be known according to the reconstructed target profile.

\subsection{Material Identification} \label{section_Material_Identification}
To ensure real-time prediction and classification of potential threat objects in the reflectarray screening system, it is desired to develop a fully analytical forward model for characterizing the complex relative permittivity of the object under detection\cite{1451581, 1451580}. To achieve this, a ray tracing based GO method is proposed to predict the received electric fields $\tilde E_n^{{\rm{rec}}}$\cite{8609050}. By sweeping the relative permittivity $ \varepsilon _r =  \varepsilon _r'  - j \varepsilon _r''$ and the object thickness $T$, $\tilde E_n^{{\rm{rec}}}( \varepsilon _r', \varepsilon _r'', T)$ are calculated and compared to the measured $E_n^{{\rm{rec}}}$ (or simulated with the full-wave method of multilevel fast multipole algorithm, MLFMA) to find the best matched magnitude and phase responses. Consequently, the corresponding estimated object thickness $\tilde T$ and relative permittivity $\tilde \varepsilon _r = \tilde \varepsilon _r'  - j\tilde \varepsilon _r''$ are obtained.

\begin{figure}[t]
\centering
\includegraphics[width=1\linewidth]{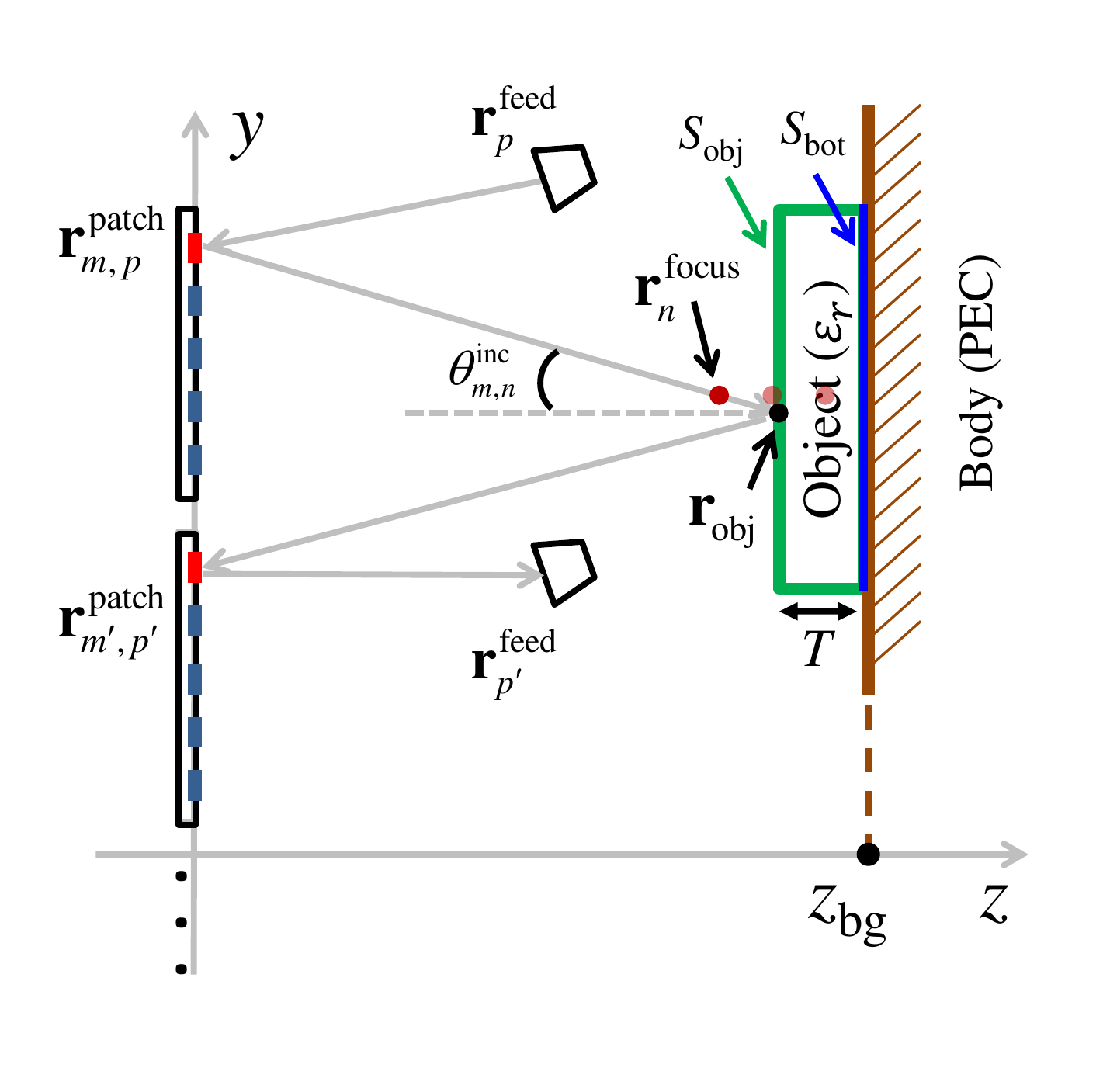}
\caption{GO model for the object material identification. All the rays, originating from the transmitting horns and terminating at the receiving horns, are considered for estimating the unknown complex relative permittivity $\varepsilon_r$ and the thickness $T$. $S_{\mathrm{obj}}$ and $S_{\mathrm{bot}}$ represent the air-object and object-body interface, respectively. ${z_{\mathrm{bg}}}$ is the range of $S_\mathrm{bot}$.}
\label{materials_characterization}
\end{figure}
Figure \ref{materials_characterization} shows the proposed forward model for a multi-reflectarray-based system. For a general analysis, the $p$-th feeding antenna at ${\bf{r}}_p^{{\rm{feed}}}$ is assumed to be active, and only the ray that originates from ${\bf{r}}_p^{{\rm{feed}}}$ and terminates at the $p'$-th feeding antenna in the receiving mode at ${\bf{r}}_{p'}^{{\rm{feed}}}$ is considered.
Letting the ray at ${\bf{r}}_p^{{\rm{feed}}}$ has an amplitude unity of $E_0$, the complex incident amplitude $\tilde E_{m,p}^{{\rm{patch}}}$ on the $m$-th patch of the $p$-th reflectarray can be computed by considering both magnitude loss and phase delay,
\begin{equation}
\tilde E_{m,p}^{{\rm{patch}}} = \frac{{{E_0}{e^{ - j{k_0}\left| {{\bf{r}}_{m,p}^{{\rm{patch}}} - {\bf{r}}_p^{{\rm{feed}}}} \right|}}}}{{\left| {{\bf{r}}_{m,p}^{{\rm{patch}}} - {\bf{r}}_p^{{\rm{feed}}}} \right|}},
\label{eq_Epatch_mp}
\end{equation}
where ${\bf{r}}_{m,p}^{{\rm{patch}}}$ is the position of the $m$-th patch of the $p$-th reflectarray.

Reflected from the $p$-th reflectarray, which is assumed to focus the CW wave at the $n$-th focusing point ${\bf{r}}_{n}^{{\rm{focus}}}$, the ray will reach the air-object interface $S_{\mathrm{obj}}$ at the point ${\bf{r}}_{{\rm{obj}}}$ using ray tracing,
\begin{equation}
{{\bf{r}}_{{\rm{obj}}}} = \frac{{{z_{\mathrm{bg}}} - T}}{{\cos \theta _{n,m}^{{\rm{inc}}}}} \frac{{{\bf{r}}_n^{{\rm{focus}}} - {\bf{r}}_{m,p}^{{\rm{patch}}}}}{{\left| {{\bf{r}}_n^{{\rm{focus}}} - {\bf{r}}_{m,p}^{{\rm{patch}}}} \right|}} + {\bf{r}}_{m,p}^{{\rm{patch}}}
\label{eq_robj}
\end{equation}
where ${z_{\mathrm{bg}}}$ is the range of the object-body interface $S_\mathrm{bot}$; $T$ is the thickness of the dielectric object; and ${{\theta _{n,m}^{{\rm{inc}}}}}$ is the incident angle. Thus, the ray amplitude at ${\bf{r}}_{{\rm{obj}}}$ can be expressed as
\begin{equation}
\tilde E_{n,m,p}^{{\rm{obj}}} =  - \frac{{\tilde E_{m,p}^{{\rm{patch}}} {e^{ - j\left[ {{k_0}\left( {\left| {{{\bf{r}}_{{\rm{obj}}}} - {\bf{r}}_{m,p}^{{\rm{patch}}}} \right|} \right) - \Delta {\psi _{n,m,p}}} \right]}}}}{{1{\rm{ + }}\left| {{{\bf{r}}_{{\rm{obj}}}} - {\bf{r}}_{m,p}^{{\rm{patch}}}} \right|/\left| {{\bf{r}}_{m,p}^{{\rm{patch}}} - {\bf{r}}_p^{{\rm{feed}}}} \right|}},
\label{eq_Eobj}
\end{equation}
where $\Delta \psi_{n,m,p}$ is the binary phase shift added on the $m$-th patch element of the $p$-th reflectarray when focusing at ${\bf{r}}_{n}^{{\rm{focus}}}$, which is defined in Eq. (\ref{eq_binary_phase_correction}).

Scattered by both the dielectric and body surfaces, considering multiple reflections within the dielectric object, the backwards ray will illuminate upon the $m'$-th patch of the $p'$-th reflectarray at ${{\bf{r}}_{m',p'}^{{\rm{patch}}}}$. Note that the subindexes $m'$ and $p'$ of ${{\bf{r}}_{m',p'}^{{\rm{patch}}}}$ can be determined using ray tracing again, which are only dependent on ${{\bf{r}}_{m,p}^{{\rm{patch}}}}$ and ${{\bf{r}}_{\mathrm{obj}}}$. Thus, the ray at ${{\bf{r}}_{m',p'}^{{\rm{patch}}}}$ will have an amplitude of $\tilde E_{n,m,p}^{{\rm{patch}}}$,
\begin{equation}
\tilde E_{p',m',n,m,p}^{{\rm{patch}}} = \frac{{\tilde E_{n,m,p}^{{\rm{obj}}} \Gamma \left( {\theta _{n,m}^{{\rm{inc}}}} \right) \cdot {e^{ - j{k_0}\left( {\left| {{\bf{r}}_{m',p'}^{{\rm{patch}}} - {{\bf{r}}_{{\rm{obj}}}}} \right|} \right)}}}}{{{\rm{1 + }}\left| {{\bf{r}}_{m',p'}^{{\rm{patch}}} - {{\bf{r}}_{{\rm{obj}}}}} \right|/\left| {{{\bf{r}}_{{\rm{obj}}}} - {\bf{r}}_{m',p'}^{{\rm{patch}}}} \right|}},
\label{eq_Epatch_nmp}
\end{equation}
where $\Gamma \left( {{\theta _{n,m}^{{\rm{inc}}}}} \right)$ is the total reflection coefficient at the air-object interface $S_{\mathrm{obj}}$.

\begin{figure}[t]
\centering
\includegraphics[width=0.8\linewidth]{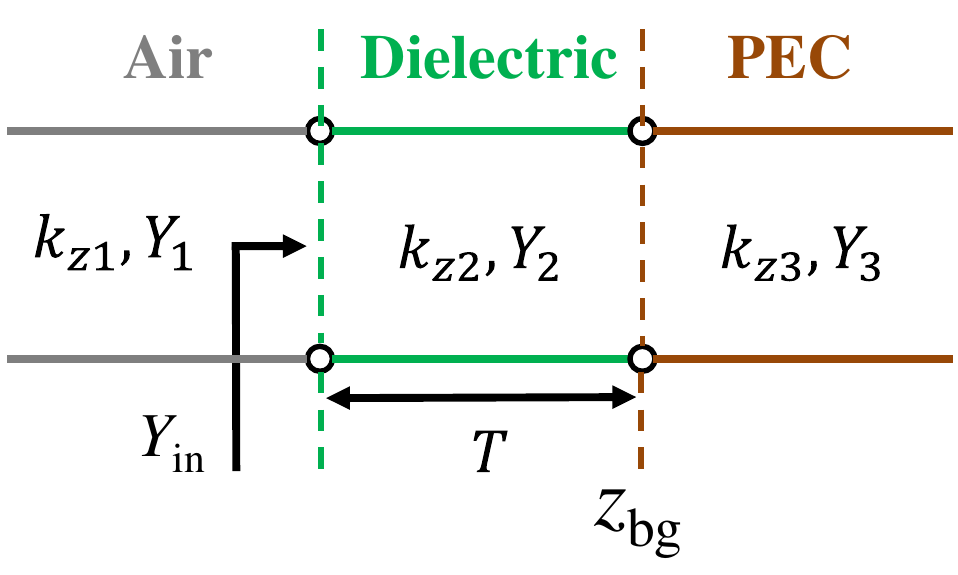}
\caption{Transmission line model to characterize the total reflection coefficient $\Gamma \left( {{\theta _{n,m}^{{\rm{inc}}}}} \right)$, where $k_{zi}$, $i \in \{1, 2,3\}$, is the wave number in the $z$ direction for the air, dielectric, and body (PEC) layers, respectively; $Y_i$ is the characteristic admittance for each layer; $Y_{\mathrm{in}}$ is the input admittance at the air-object interface; $T$ is the thickness of the dielectric object; and ${z_{\mathrm{bg}}}$ is the range of the object-body interface $S_\mathrm{bot}$.}
\label{transmission_line_model}
\end{figure}
To characterize the total reflection coefficient $\Gamma \left( {{\theta _{n,m}^{{\rm{inc}}}}} \right)$, the transmission line model is introduced in Fig. \ref{transmission_line_model}. It can be an effective model because (1) the electrically large reflectarray is capable of focusing the incident wave into a tiny spot in the RoI, which is much smaller compared to the object dimensions in the transverse ($x$-$y$) plane so that the edge diffraction effect from the object can be circumvented; and (2) the dielectric object is already in the far-field region of the patch elements in the reflectarrays such that it is suitable to use the plane-wave incidence approximation in the RoI.

Denoting $k_{zi}$, $i \in \{1, 2,3\}$, as the wave number in the $z$-direction for the air, dielectric, and body (PEC) layer, respectively, the corresponding characteristic admittance $Y_i$ for TE and TM mode can be calculated as follows:
\begin{eqnarray}
{Y_i} = \left\{ \begin{array}{l}
\frac{{\omega {\varepsilon _{ri}}{\varepsilon _0}}}{{{k_{zi}}}}{\rm{ ,\quad for \quad TM \quad mode}}\\
\frac{{{k_{zi}}}}{{\omega \mu_0 }}{\rm{,\quad for \quad TE \quad mode}}
\end{array} \right.
\label{eq_characteristic_admittance}
\end{eqnarray}
where ${k_{zi}} = \sqrt {k_0^2{\varepsilon _{ri}} - k_x^2}$; $k_0^2 = {\omega ^2}\mu {\varepsilon _0}$; ${k_x} = {k_0}\sin {{\theta _{m,n}^{{\rm{inc}}}}}$; $\varepsilon _0$ and $\mu_0$ are the vacuum permittivity and permeability, respectively; and $\varepsilon _{r1}$, $\varepsilon _{r2}$, and $\varepsilon _{r3}$ are the relative permittivities of the air, dielectric object, and PEC, respectively. The input admittance $Y_{\mathrm{in}}$ at the air-dielectric interface can be wrote as
\begin{eqnarray}
{Y_{{\rm{in}}}} = {Y_2}\frac{{{Y_3} + j{Y_2}\tan {k_{z2}}T}}{{{Y_2} + j{Y_3}\tan {k_{z2}}T}}.
\label{eq_input_admittance1}
\end{eqnarray}
Accordingly, the total reflection coefficient $\Gamma \left( {{\theta _{n,m}^{{\rm{inc}}}}} \right)$ is expressed as
\begin{eqnarray}
\Gamma \left( {{\theta _{n,m}^{{\rm{inc}}}}} \right)  = \frac{{{Y_1} - {Y_{{\rm{in}}}}}}{{{Y_1} + {Y_{{\rm{in}}}}}}.
\label{eq_reflection_coefficient}
\end{eqnarray}

With the confocal setup of the reflectarrays, the ray is refocused by the $p'$-th reflectarray and directed towards the corresponding $p'$-th feeding antenna located at ${{\bf{r}}_{p'}^{{\rm{feed}}}}$. The received complex amplitude $\tilde E_{n,m,p}^{{\rm{rec}}}$ at ${{\bf{r}}_{p'}^{{\rm{feed}}}}$ is
\begin{equation}
\begin{array}{l}
\begin{aligned}
\tilde E_{p',m',n,m,p}^{{\rm{rec}}} = & - \tilde E_{p',m',n,m,p}^{{\rm{patch}}} \times \\
&\frac{{{e^{ - j\left[ {{k_0}\left( {\left| {{\bf{r}}_{p'}^{{\rm{feed}}} - {\bf{r}}_{m',p'}^{{\rm{patch}}}} \right|} \right) - \Delta {\psi _{n,m',p'}}} \right]}}}}{{1 + \left| {{\bf{r}}_{p'}^{{\rm{feed}}} - {\bf{r}}_{m',p'}^{{\rm{patch}}}} \right|/\left| {{\bf{r}}_{m',p'}^{{\rm{patch}}} - {\bf{r}}_n^{{\rm{focus}}}} \right|}},
\end{aligned}
\end{array}
\label{eq_estimate_Erec}
\end{equation}
where $\Delta \psi_{n,m',p'}$ is the binary phase shift added on the $m'$-th patch element of the $p'$-th reflectarray when refocusing from ${\bf{r}}_{n}^{{\rm{focus}}}$.

Finally, the total received complex amplitude when focusing at ${\bf{r}}_{n}^{{\rm{focus}}}$ can be calculated by a summation:
\begin{eqnarray}
\tilde E_n^{{\rm{rec}}} = \sum\limits_{p' = 1}^P {\sum\limits_{m' = 1}^M {\sum\limits_{m = 1}^M {\sum\limits_{p = 1}^P {\tilde E_{p',m',n,m,p}^{{\rm{rec}}}} } } } ,
\label{eq_analytical_Erec}
\end{eqnarray}
where $P$ is the total number of FARAs; and $M$ is the total number of patch elements at each reflectarray.

By sweeping $\varepsilon_r'$ and $\varepsilon_r''$ of the relative permittivity $\varepsilon_r  = \varepsilon_r' -j \varepsilon_r''$, and the thickness $T$, the GO predicted received amplitude ${{\tilde E_n^{{\rm{rec}}}}}\left( {\varepsilon _r',\varepsilon _r'',T} \right)$, $n\in[1,N]$, $N$ being the total number of focusing points used in the estimation, are calculated, and compared to the measured ${{E_n^{{\rm{rec}}}}}$ to find the best match by means of minimizing the error function $f\left( {{\varepsilon _r'},\varepsilon _r'',T} \right)$, namely
\begin{equation}
\left.
\begin{array}{c}
\left\{ {{\tilde \varepsilon }_r'},{\tilde \varepsilon }_r'', \tilde T \right\} = \arg \mathop {\min }\limits_{{{ \varepsilon }_r'}, {\varepsilon _r''}, T} \left\{ {f\left( {\varepsilon _r',\varepsilon _r'',T} \right)} \right\} \\
\mathrm{s.t.} \quad f\left( {{\varepsilon _r'},{\varepsilon _r''},T} \right) = \sum\limits_{n = 1}^N {\left| {\frac{{\tilde E_n^{{\rm{rec}}}}}{{\tilde E_0^{{\rm{rec}}}}} - \frac{{E_n^{{\rm{rec}}}}}{{E_0^{{\rm{rec}}}}}} \right|},
\end{array}
\right.
\label{eq_estimation_eps}
\end{equation}
where ${{\tilde \varepsilon }_r'}$, ${{\tilde \varepsilon }_r''}$, and ${\tilde T}$ are the estimated object dielectric constant, loss factor, and thickness, respectively; and ${\tilde E_0^{{\rm{rec}}}}$ and ${E_0^{{\rm{rec}}}}$ are the calibration amplitudes to normalize the predicted and measured received fields, respectively, which are independent on the dielectric object under detection, and can be measured by focusing the incident wave at a reference plane, located at a range different from or the same as $z_\mathrm{bg}$.
% For the three undetermined parameters, namely $T$, $\varepsilon_r'$, and $\varepsilon_r''$, $N = 3$ is considered for the estimation in the following simulations and experiment.
% Note that, with larger values of $N$, it is possible to achieve a higher estimation accuracy; however, it suffers from a heavier computational burden, as well as a longer estimation time.

\section{Primary Results} \label{section_results}
\begin{figure}[t]
\centering
\includegraphics[width=\linewidth]{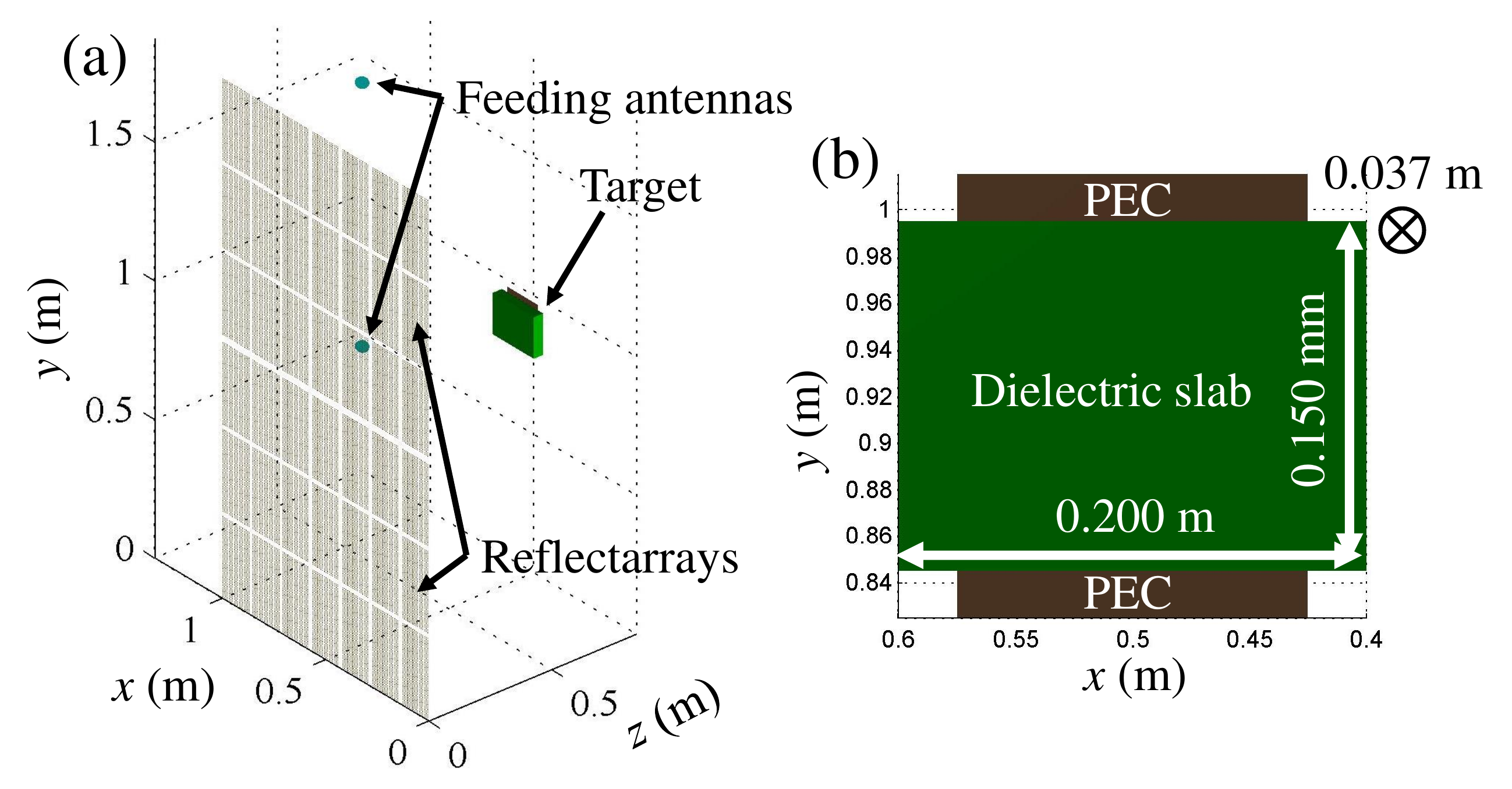}
\caption{3D setup for simulation and experiment: (a) two reflectarrays are vertically stacked and fed by two $10$ dB standard gain pyramidal horn antennas, operated at $24.16$ GHz; and (b) the dielectric object is attached to the center of the PEC plate.}
\label{simulation_setup}
\end{figure}
To validate the proposed method for object profiles reconstruction and materials identification, both simulated and experimental examples are examined. The simulation setup is the same as that in the experiment, which is shown in Fig. \ref{simulation_setup}. Each reflectarray has a side length of $1000$ mm. The centers of the top and the bottom reflectarray are located at $[500, 1413, 0]$ mm and $[500, 461, 0]$ mm, respectively. Two identical $10$ dB standard gain pyramidal horn antennas are placed at $[1313, 1413, 830]$ mm and $[1313, 461, 830]$ mm, respectively, facing the corresponding centers of the reflectarrays. The horn antennas are operated at the single frequency $24.16$ GHz, corresponding to a wavelength of $\lambda_0 \approx 12.4$ mm. The dielectric object is a slab that is attached to the center $[500, 920, 800]$ mm of a steel plate (${z_{\mathrm{bg}}} =  800$ mm). The slab has the dimensions of $200$ mm, $150$ mm, and $37$ mm in the $x$-, $y$-, and $z$-axis, respectively. Note that the object thickness can be varied in different simulated and experimental cases.

\subsection{Simulation Results}
\begin{figure}[t]
\centering
\includegraphics[width=0.8\linewidth]{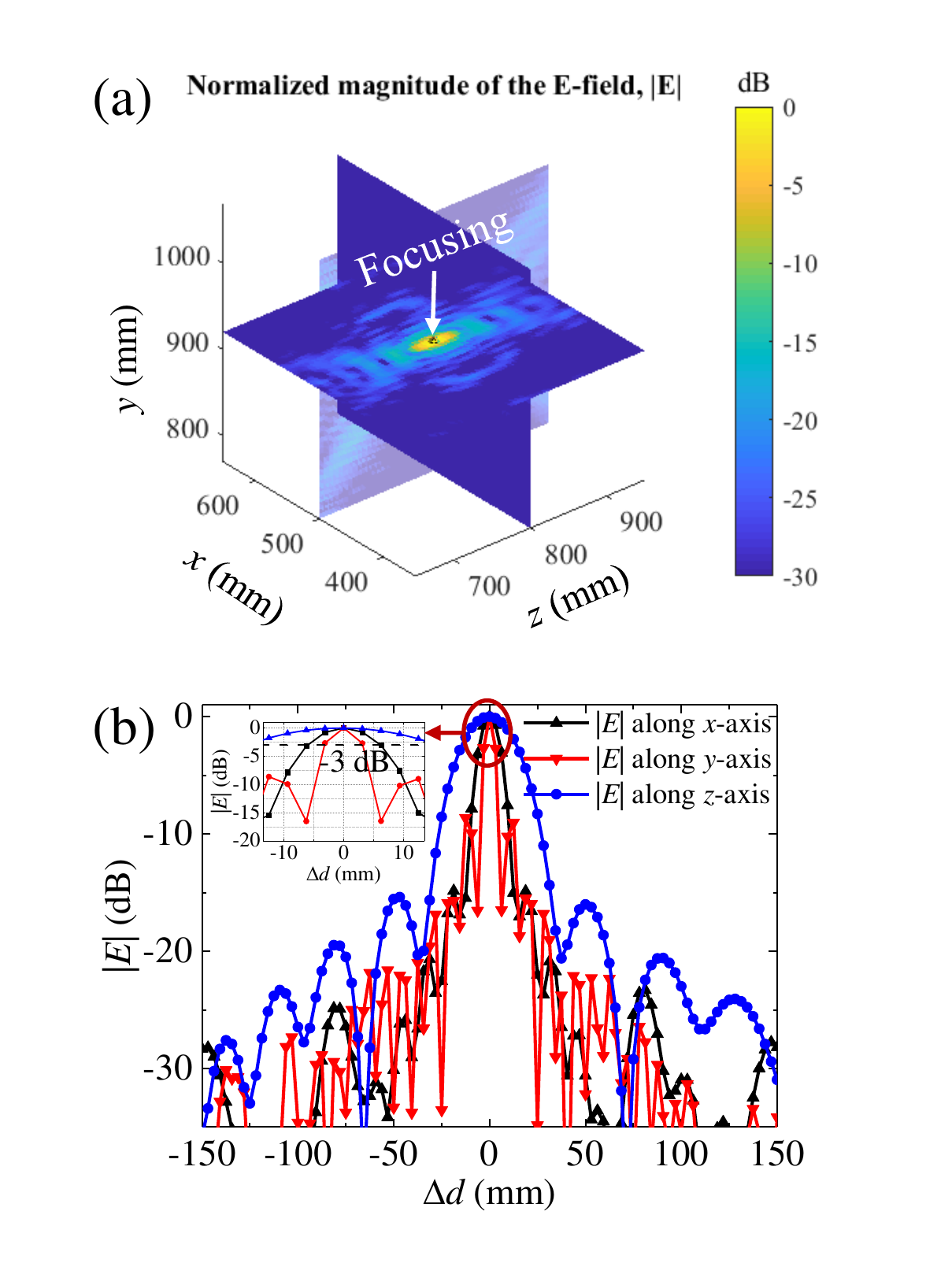}
\caption{PSF of the reflectarray imaging system when the focusing is at $[500, 920, 800]$ mm in free-space. (a) and (b) show the 3D and 2D radiation patterns, respectively. $\Delta d$ is the distance between the focusing point and the observation point.}
\label{psf}
\end{figure}
First of all, it is important to examine the focusing quality of the reflectarrays based on the point spread function (PSF)\cite{JMI:JMI3168}. By setting the focusing point at $[500, 920, 800]$ mm in free-space, the PSF is calculated. Figure \ref{psf}(a) and \ref{psf}(b) show the 3D and 2D radiation patterns, respectively. As anticipated, a sharp focusing spot, namely high imaging resolution, is achieved. The $3$-dB width of the focusing spot is near ${\lambda_0}/2$ along the $x-$ and $y-$axis, and ${\lambda_0}$ along the $z-$axis.

In order to obtain a high calculation efficiency while retaining an acceptable imaging accuracy, the $3^{\mathrm{rd}}$-order PO is used to calculate the electric and magnetic currents on the dielectric object surface, namely applying Eq. (\ref{eq_JM_obj_multi}) with $K = 3$.

First, two pure dielectric (lossless) objects are considered in the simulation. Object$-1$ has a thickness of $T_1 = 20$ mm and relative permittivity of $\varepsilon_{r1} = 8.0-j0.0$; while Object$-2$ has a thickness of $T_2 = 40$ mm and relative permittivity of $\varepsilon_{r2} = 2.0-j0.0$. These two objects, satisfying ${{T_1}\sqrt {{{\varepsilon_{r1} '}}} = {T_2}\sqrt {{{\varepsilon_{r2} '}}} }$, are selected to verify that the proposed method is able to solve the phase-shift ambiguity\cite{Trabelsi2000Phase} without loss of generality.

\begin{figure}[t]
\centering
\includegraphics[width=0.95\linewidth]{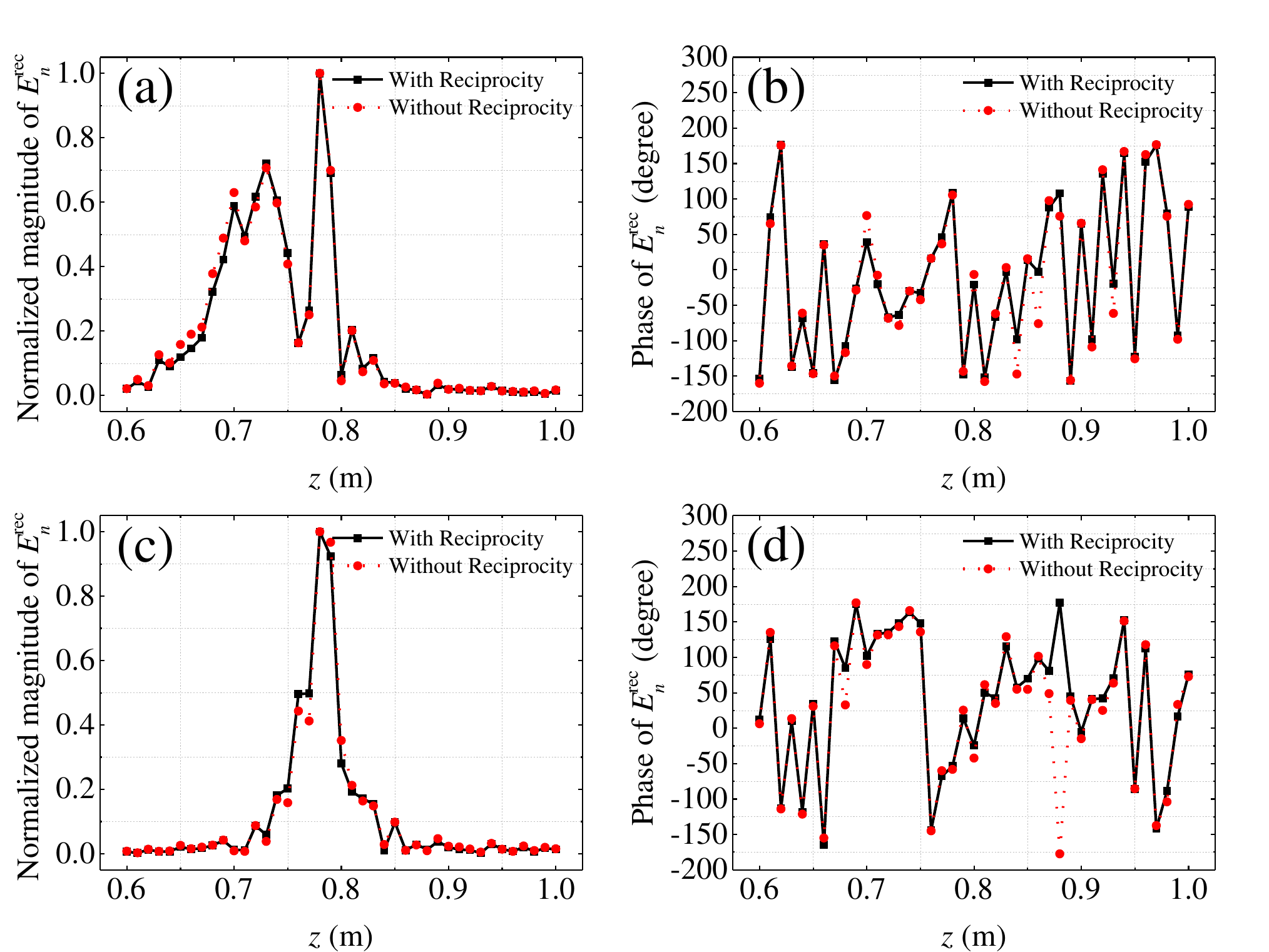}
\caption{Received field, with and without applying the reciprocity theorem, for the simulation setup described in Fig. \ref{simulation_setup}. The focusing point is uniformly swept from $[500, 920, 600]$ mm to $[500, 920, 1000]$ mm along the $z$-axis. (a) and (b) are the calculated magnitude and phase distribution, respectively, for Object$-1$; and, similarly, (c) and (d) are the calculated magnitude and phase distribution, respectively, for Object$-2$.}
\label{reciprocity_theorem}
\end{figure}

In addition, the general reciprocity theorem described in Eq. (\ref{eq_reciprocity_theorem}) are applied throughout the simulations to improve the computation efficiency. Its effectiveness is validated in Fig. \ref{reciprocity_theorem}, where the received fields, with and without the reciprocity theorem, are in a good agreement for both Object$-1$ and Object$-2$. Figure \ref{reciprocity_theorem} also verifies that the radiation pattern of the reflectarray in the near-field is range-dependent so that the magnitude and phase responses of the received fields for the two ambiguous objects are distinguished, which shows the effectiveness to discriminate these two objects using the algorithm derived in Eq. (\ref{eq_estimation_eps}).

\begin{figure}[t]
\centering
\includegraphics[width=\linewidth]{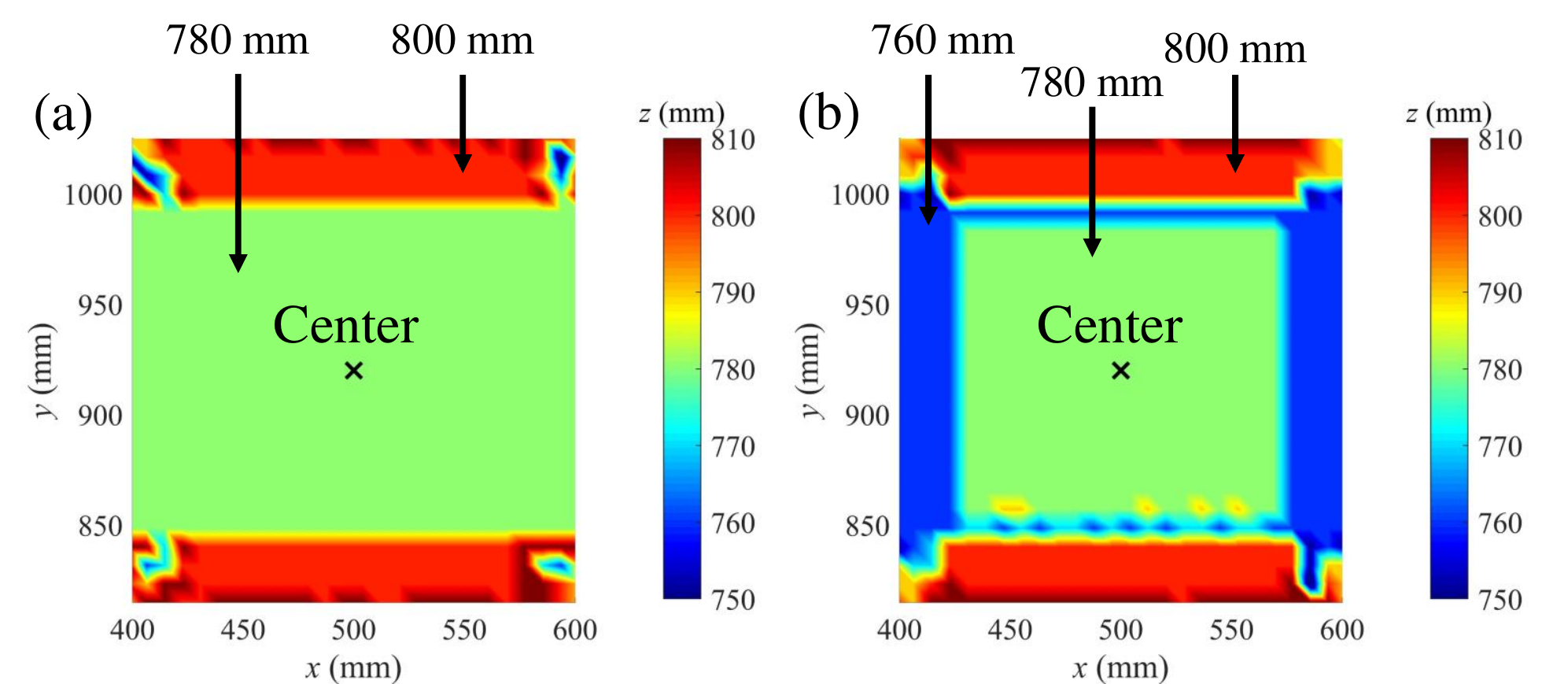}
\caption{(a) PO simulated target profile for Object$-1$ ($\varepsilon_{r1} = 8.0 - j0.0$ and $T_1 = 20$ mm), where the imaged thickness and profile center are $20$ mm and $[x_{\mathrm{c}1}, y_{\mathrm{c}1}, z_{\mathrm{c}1}] = [500, 920, 780]$ mm, respectively; and (b) PO simulated target profile for Object$-2$ ($\varepsilon_{r2} = 2.0 - j0.0$ and $T_2 = 40$ mm), where the imaged thickness and profile center are $20$ mm and $[x_{\mathrm{c}2}, y_{\mathrm{c}2}, z_{\mathrm{c}2}] = [500, 920, 780]$ mm, respectively.}
\label{simulation_profile_lossless}
\end{figure}
Figure \ref{simulation_profile_lossless}(a) and \ref{simulation_profile_lossless}(b) show the PO simulated target profiles for Object$-1$ and Object$-2$, respectively, which give the accurate widths, in the $x$-axis, and heights, in the $y$-axis, for both Object$-1$ and Object$-2$. As it is also seen, for Object$-1$ with an actual thickness of $20$ mm, the imaged profile center and thickness are $[x_{\mathrm{c}1}, y_{\mathrm{c}1}, z_{\mathrm{c}1}] = [500, 920, 780]$ mm and $800 - 780 = 20$ mm, respectively; however, for Object$-2$ that has an actual thickness of $40$ mm, the imaged profile center and thickness are also $[x_{\mathrm{c}2}, y_{\mathrm{c}2}, z_{\mathrm{c}2}] = [500, 920, 780]$ mm and $800 - 780 = 20$ mm, respectively. This phenomenon is attributed to the multiple reflections inside the dielectric object so that the focusing position, corresponding to the maximum magnitude of the received field along the $z$-axis, is achieved under the front surface of the dielectric object.

\begin{figure}[t]
\centering
\includegraphics[width=\linewidth]{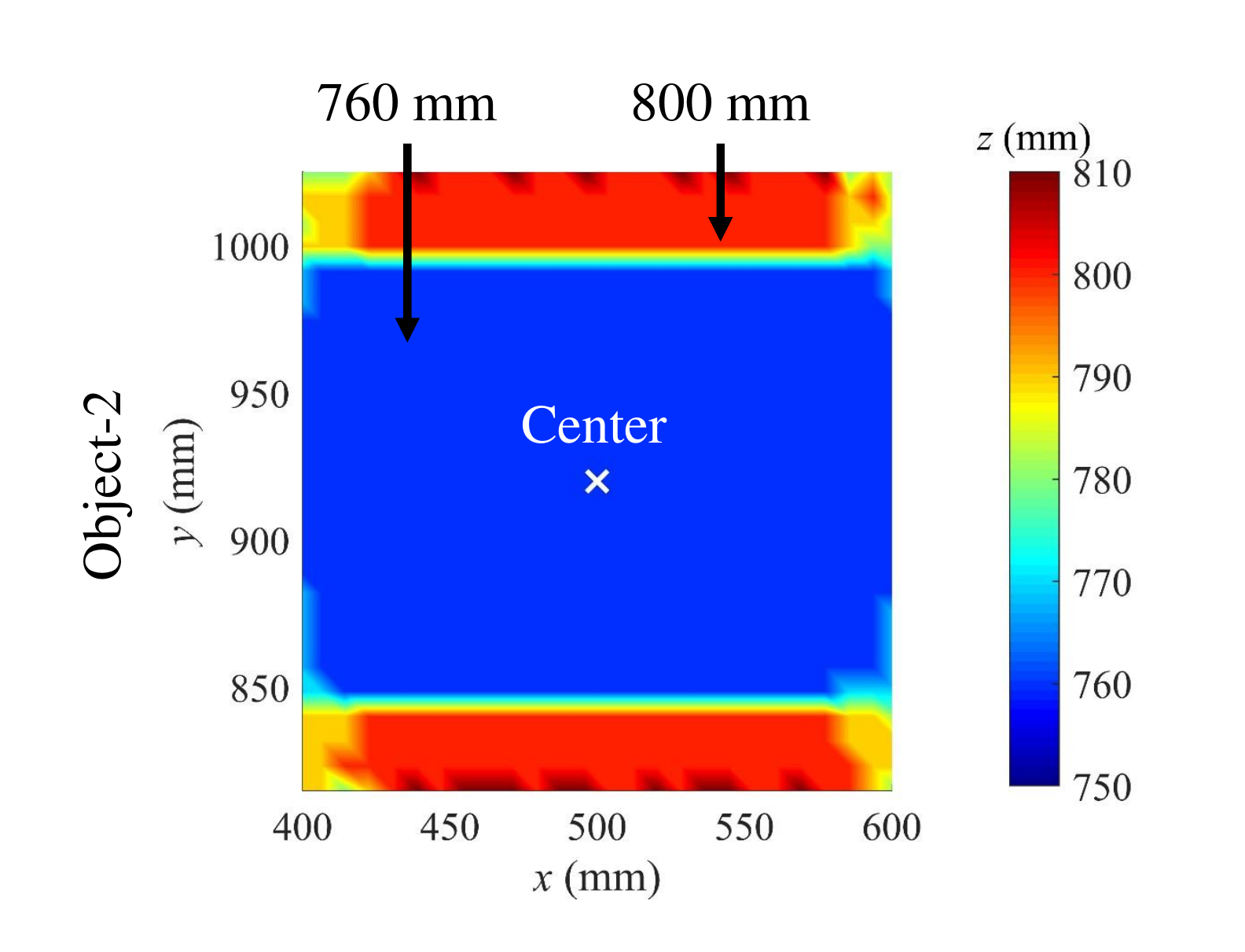}
\caption{PO simulated target profile for Object$-3$ ($\varepsilon_{r3} = 4.0 - j0.2$ and $T_3 = 40$ mm), where the imaged thickness and profile center are $40$ mm and $[x_{\mathrm{c}3}, y_{\mathrm{c}3}, z_{\mathrm{c}3}] = [500, 920, 760]$ mm, respectively.}
\label{simulation_profile_lossy}
\end{figure}
When it occurs to a lossy Object$-3$, assuming it has a thickness of $T_3 = 40$ mm and complex relative permittivity of $\varepsilon_{r3} = 4.0-j0.2$, the magnitude of the multiple reflected waves inside the lossy dielectric slab are considerably degenerated due to the large propagation loss, $\epsilon_r'' = 0.2$. The imaging processing of Object$-3$ can be similar to that of a metallic object, where the $1^{\mathrm{st}}$-order PO method can be sufficient to obtain a quite accurate target profile, and the maximum received field along $z$-axis can be achieved when the focusing position is near the front surface of the dielectric object. Figure \ref{simulation_profile_lossy} shows the PO simulated profile of Object$-3$, where the imaged profile center and thickness are $[x_{\mathrm{c}3}, y_{\mathrm{c}3}, z_{\mathrm{c}3}] = [500, 920, 760]$ mm and $800 \; \mathrm{mm} - 760 \; \mathrm{mm} = 40 \; \mathrm{mm}$, respectively.

Considering all aforementioned simulation cases, for an object made of unknown material, a further estimation algorithm, in addition to the profile reconstruction, is required to estimate not only the object permittivity but also a more precise object thickness.

Applying the material identification method derived in Section \ref{section_Material_Identification}, $\varepsilon_r'$ and $\varepsilon_r''$ are swept from $2.0$ to $10.0$, and from $0$ to $0.5$, respectively. This sweep range covers most common threat materials, such as narcotics, explosives, and other types of contrabands \cite{Watters1995Microwave, HUSSEIN1998Review}. The dielectric slab thickness $T$ is swept from $0$ mm to $60$ mm. By determining the imaged profile center $[x_{\mathrm{c}}, y_{\mathrm{c}}, z_{\mathrm{c}}]$ for each object, the equally spaced focusing points can be selected along the range ($z$-axis),
\begin{equation}
{\bf{r}}_n^{{\rm{focus}}} = \left[ {{x_{\rm{c}}},{y_{\rm{c}}},{z_{\rm{c}}} + \Delta z\left( {n - \frac{{N + 1}}{2}} \right)} \right],n \in \left[ 1, N \right],
\label{eq_position_focusingN}
\end{equation}
where $\Delta z = 10$ mm is the range resolution of the RoI, and total $N = 3$ focusing points are considered for estimating object thickness $\tilde T$ and complex relative permittivity ${{\tilde \varepsilon }_r} = {{\tilde \varepsilon }_r'} -j {\tilde \varepsilon }_r''$. The importance for selecting those focusing points is that they always correspond to higher receiving magnitudes compared to the other focusing points along the $z-$axis; therefore, a higher receiving signal-to-noise ratio (SNR) can be achieved for a better estimation accuracy.

\begin{figure}[t]
\centering
\includegraphics[width=\linewidth]{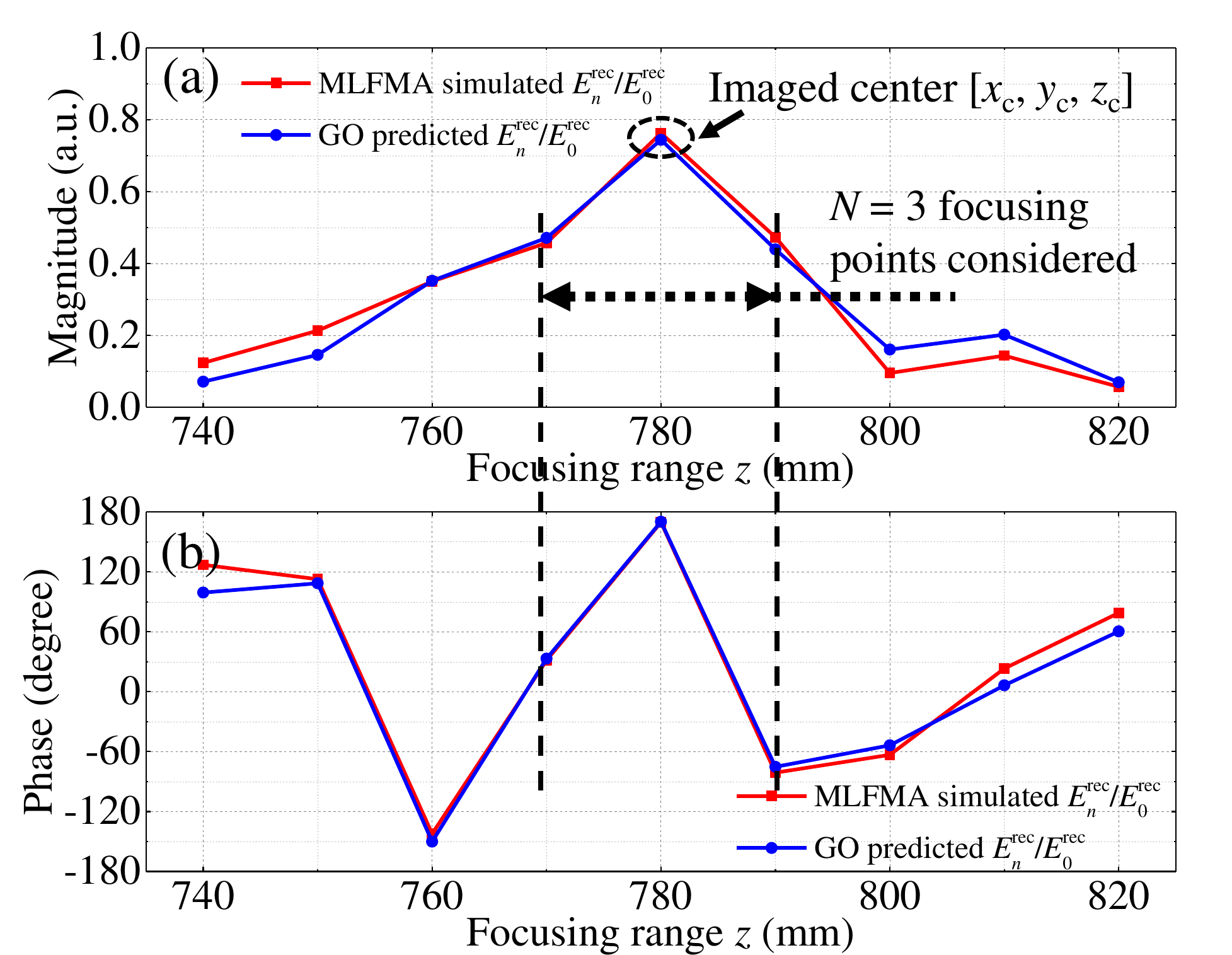}
\caption{Best matched magnitude (a) and phase (b) response for Object-$1$, when the minimum error $\mathrm{min}\left\{ f({{\varepsilon _r'},\varepsilon _r'',T}) \right\}$ is achieved at ${{\tilde \varepsilon }_r'} = 8.0$, ${\tilde \varepsilon }_r''=0.0$, and $\tilde T = 20$ mm.}
\label{best_match}
\end{figure}
\begin{figure}[t]
\centering
\includegraphics[width=\linewidth]{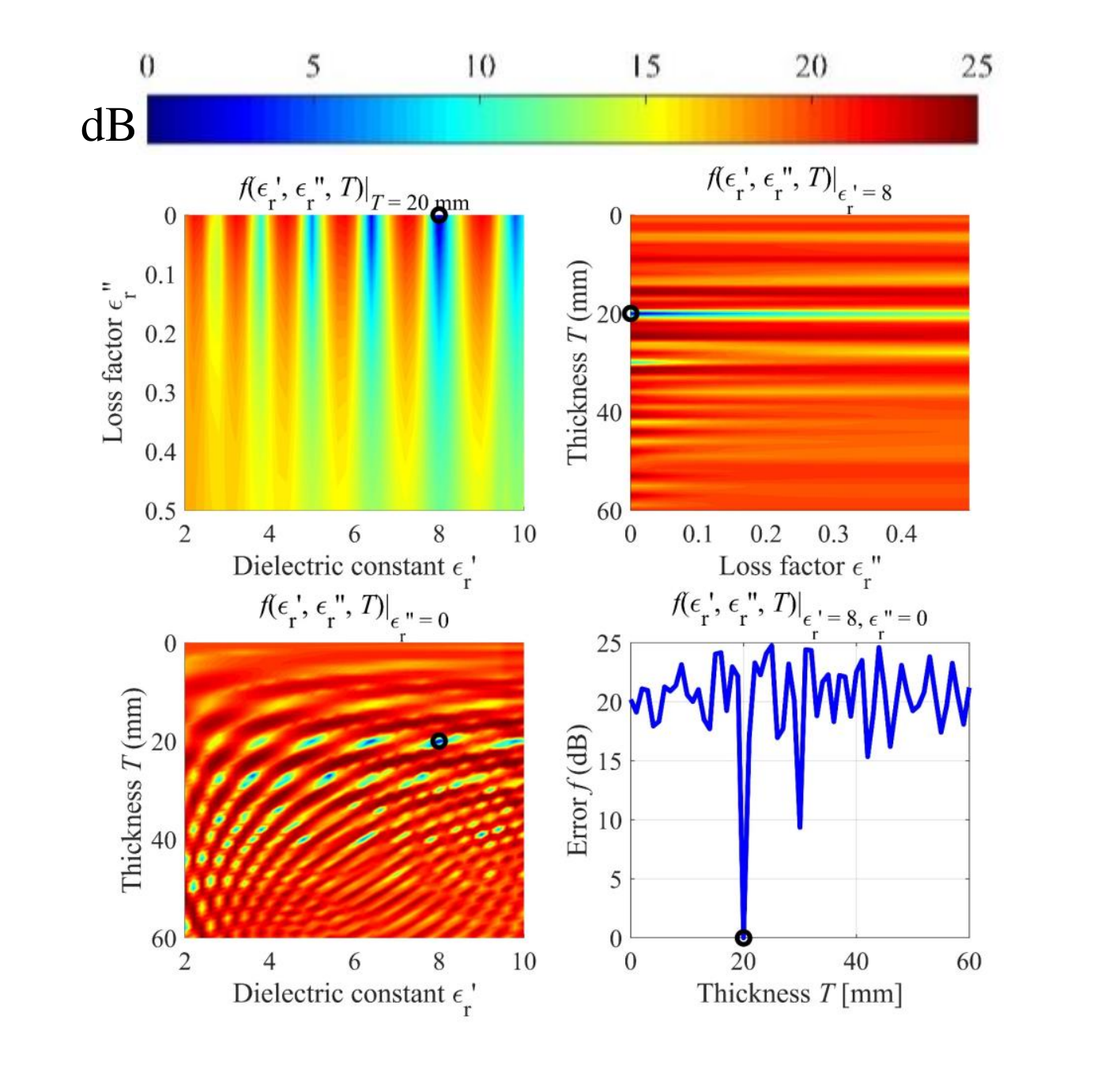}
\caption{Error distribution $f({{\varepsilon _r'},\varepsilon _r'',T})$ for Object-$1$, obtained by sweeping $\varepsilon _r'$, $\varepsilon _r''$, and $T$, where the achieved minimum error is denoted by a circle.}
\label{error_fun_obj1}
\end{figure}
\begin{figure}[t]
\centering
\includegraphics[width=\linewidth]{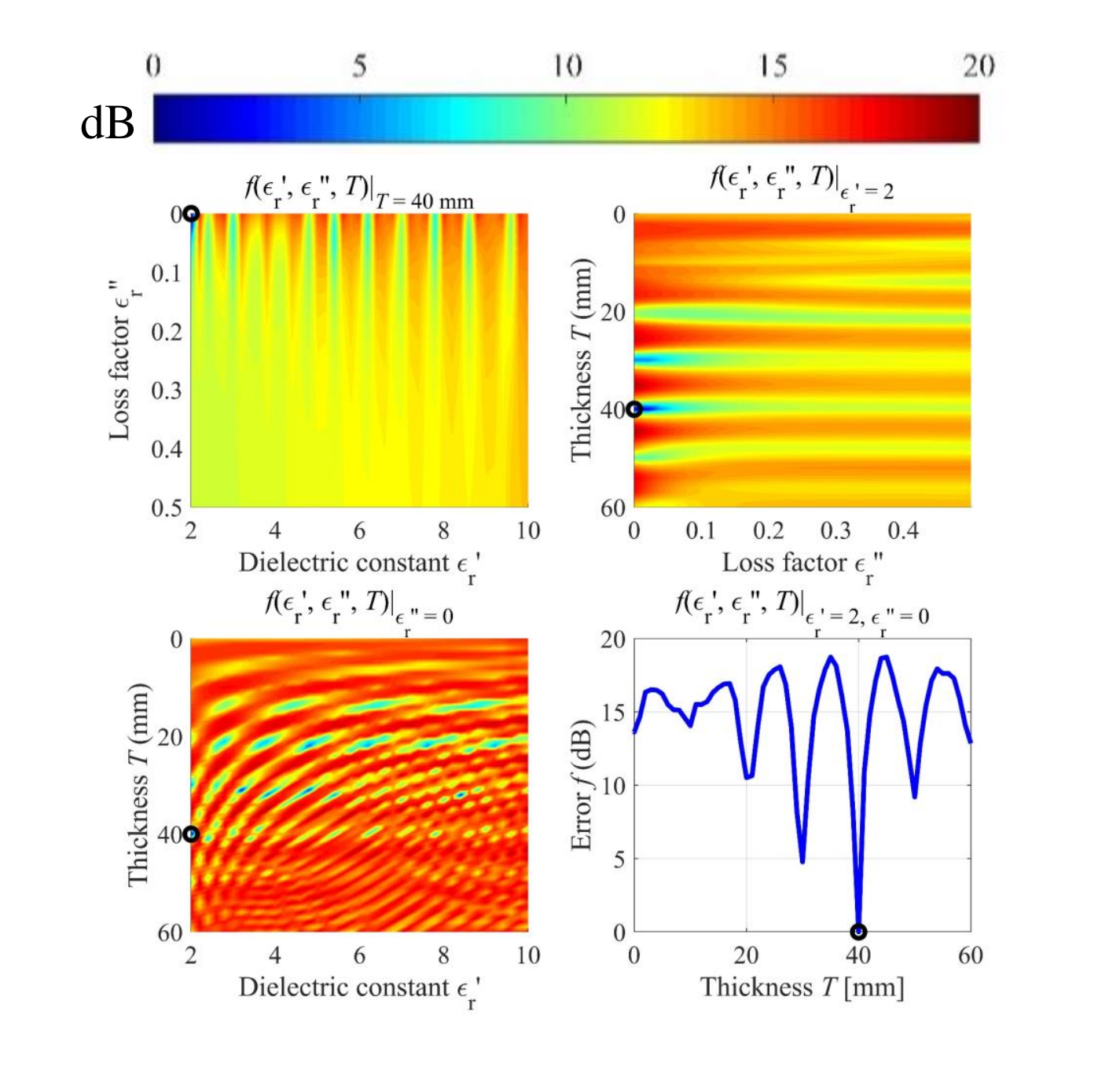}
\caption{Error distribution $f({{\varepsilon _r'},\varepsilon _r'',T})$ for Object-$2$, obtained by sweeping $\varepsilon _r'$, $\varepsilon _r''$, and $T$, where the achieved minimum error is denoted by a circle.}
\label{error_fun_obj2}
\end{figure}
\begin{figure}[t]
\centering
\includegraphics[width=\linewidth]{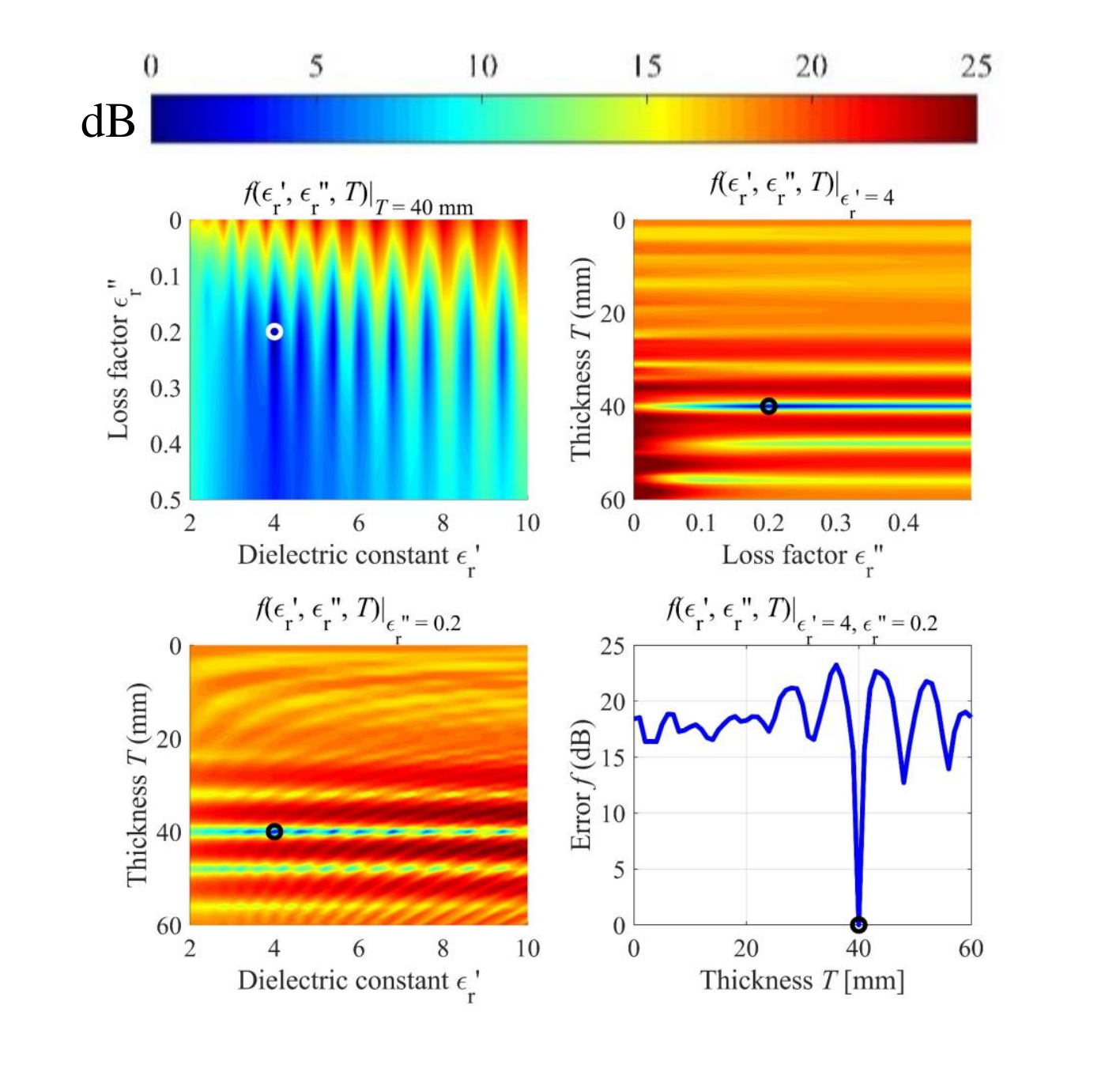}
\caption{Error distribution $f({{\varepsilon _r'},\varepsilon _r'',T})$ for Object-$3$, obtained by sweeping $\varepsilon _r'$, $\varepsilon _r''$, and $T$, where the achieved minimum error is denoted by a circle.}
\label{error_fun_obj3}
\end{figure}

 As it is shown in Fig. \ref{best_match}, the best matched magnitude (a) and phase (b) responses, corresponding to the minimum error $\mathrm{min}\left\{ f({{\varepsilon _r'},\varepsilon _r'',T}) \right\}$, for Object-$1$ are found at ${{\tilde \varepsilon }_r'} = 8.0$, ${\tilde \varepsilon }_r''=0.0$, and $\tilde T = 20$ mm. These estimated parameters are the same as the actual object thickness and relative permittivity. Note that, although more than three focusing points are presented in this figure for the purpose of verifying the accuracy of the GO predicted field, only the highlighted middle three points are utilized for the estimation. The error distribution $f({{\varepsilon _r'},\varepsilon _r'',T})$, obtained when sweeping $\varepsilon _r'$, $\varepsilon _r''$, and $T$, are shown in Fig. \ref{error_fun_obj1}, where the achieved minimum error is denoted by a circle.
The same estimation process is performed for Object-$2$, and the obtained error distribution is shown in Fig. \ref{error_fun_obj2}.
As it is seen, the minimum error $\mathrm{min}\left\{ f({{\varepsilon _r'},\varepsilon _r'',T}) \right\}$ also converges to the actual object parameters $\tilde \varepsilon_r' = 2.0$, $\tilde \varepsilon_r'' = 0.0$, and $\tilde T = 40$ mm. Therefore, these two ambiguous objects are distinguishable from each other using the proposed estimation method.
The error distribution $f({{\varepsilon _r'},\varepsilon _r'',T})$ for the lossy Object-$3$ are shown in Fig. \ref{error_fun_obj3}, where the minimum error is achieved at $\tilde \varepsilon_r' = 4.0$, $\tilde \varepsilon_r'' = 0.2$, and $\tilde T = 40$ mm, converging to the
actual object parameters again.
Accordingly, aforementioned simulation examples and results verify the proposed imaging scheme that is able to not only image the profiles but also effectively retrieve the complex relative permittivities of the dielectric objects, where a precise object thicknesses can be estimated.

\subsection{Experimental Results}
\begin{figure}[t]
\centering
\includegraphics[width=\linewidth]{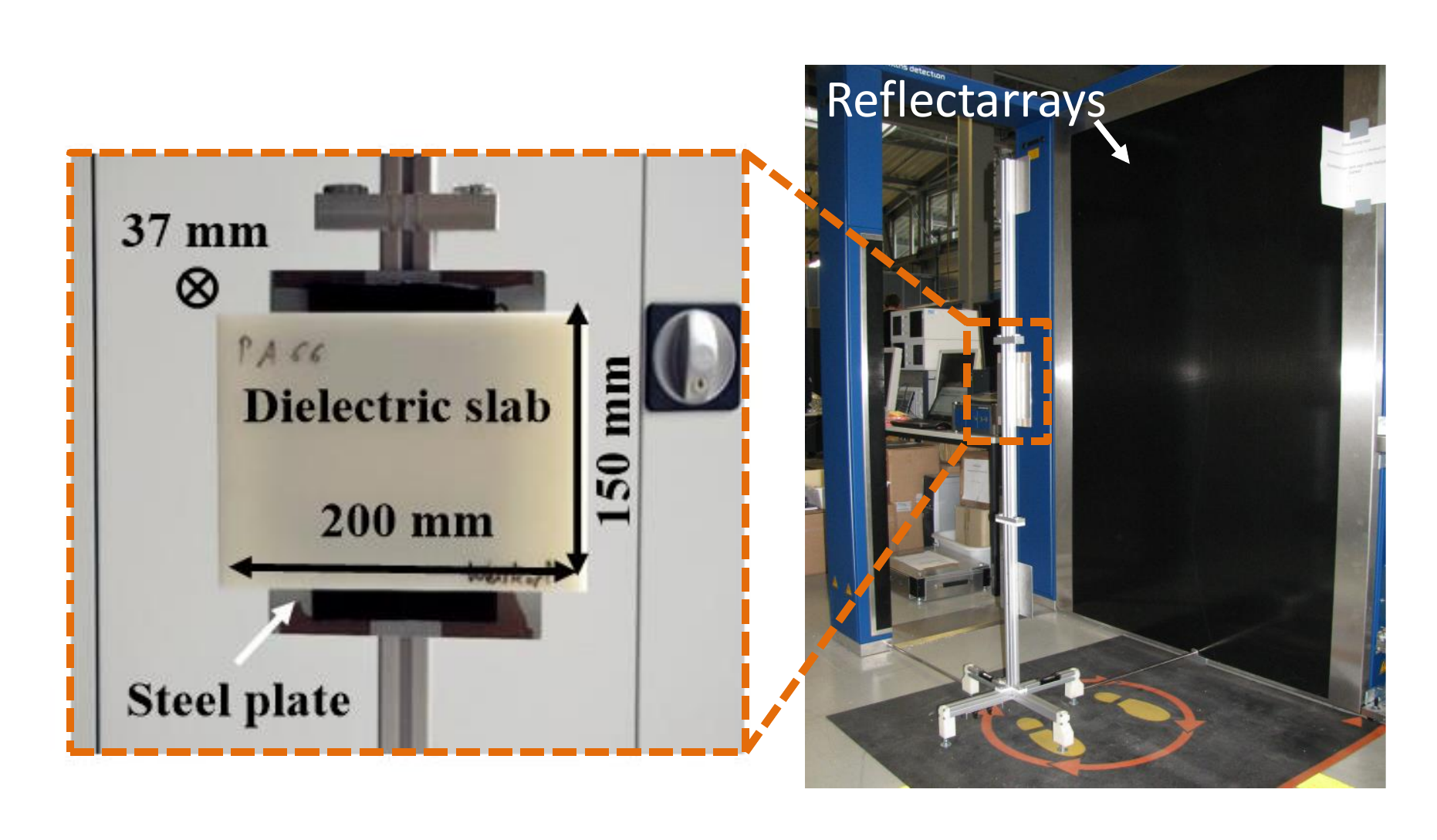}
\caption{Experimental setup to detect a dielectric object. The material of the dielectric is made of polyamide-6, 6 (PA66). The size of the dielectric is $200$ mm $\times$ $150$ mm $\times$ $37$ mm, which is attached to the center of the steel plate.}
\label{experiment_setup}
\end{figure}

The actual experiment setup to detect a threat object is given in Fig. \ref{experiment_setup}. Denote the dielectric object as Object-$4$. It is a dielectric slab made of polyamide-6, 6 (PA66), which has a relative dielectric constant of $2.8\sim3.1$ and a low loss tangent of $\tan \sigma  < 0.01$ at \textit{K}-band \cite{von1954dielectric}. The size of the dielectric is $200$ mm $\times$ $150$ mm $\times$ $37$ mm. The dielectric slab is attached to the center of a steel plate by a 1.0 mm ($<{\lambda_0}/10$) Velcro layer. To predict the received amplitude $\tilde E_n^{{\rm{rec}}}({{\varepsilon _r'},\varepsilon _r'',T})$ using GO forward model described in Fig. \ref{materials_characterization}, the Velcro layer is approximated to be an layer of air with the same thickness $1.0$ mm. Therefore, the transmission line model in Fig. \ref{transmission_line_model} for calculating the total reflection coefficient $\Gamma \left( {{\theta _{n,m}^{{\rm{inc}}}}} \right)$ is modified, which includes four cascaded layers, namely air-dielectric-air-PEC.

\begin{figure}[t]
\centering
\includegraphics[width=\linewidth]{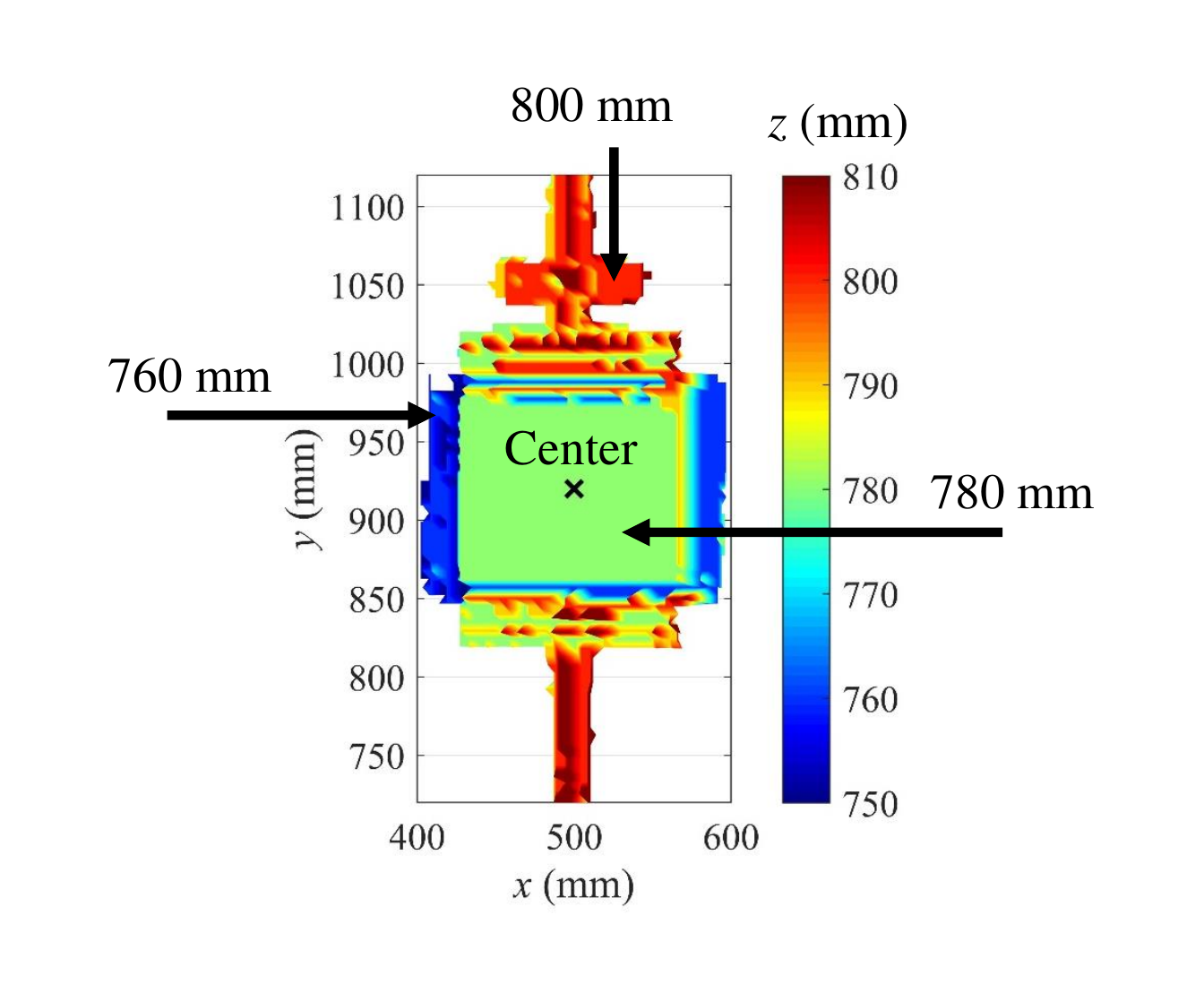}
\caption{Experimentally reconstructed target profile, where the imaged thickness and profile center are $20$ mm and $[x_{\mathrm{c}4}, y_{\mathrm{c}4}, z_{\mathrm{c}4}] = [500, 920, 780]$ mm, respectively.}
\label{experiment_profile2}
\end{figure}

Figure \ref{experiment_profile2} shows the experimentally reconstructed target profile, which has an imaged thickness of $20$ mm that is much smaller than the actual thickness of $37$ mm. The imaged profile center is at $[500,920,780]$ mm. The reason for this is that the strong multiple reflections within the dielectric slab make the focusing position, corresponding to the maximum magnitude of the received field along the $z$-axis, achieved under the front surface of the dielectric object. This explanation is the same as that for the simulation results in Fig. \ref{simulation_profile_lossless}(b).

\begin{figure}[t]
\centering
\includegraphics[width=\linewidth]{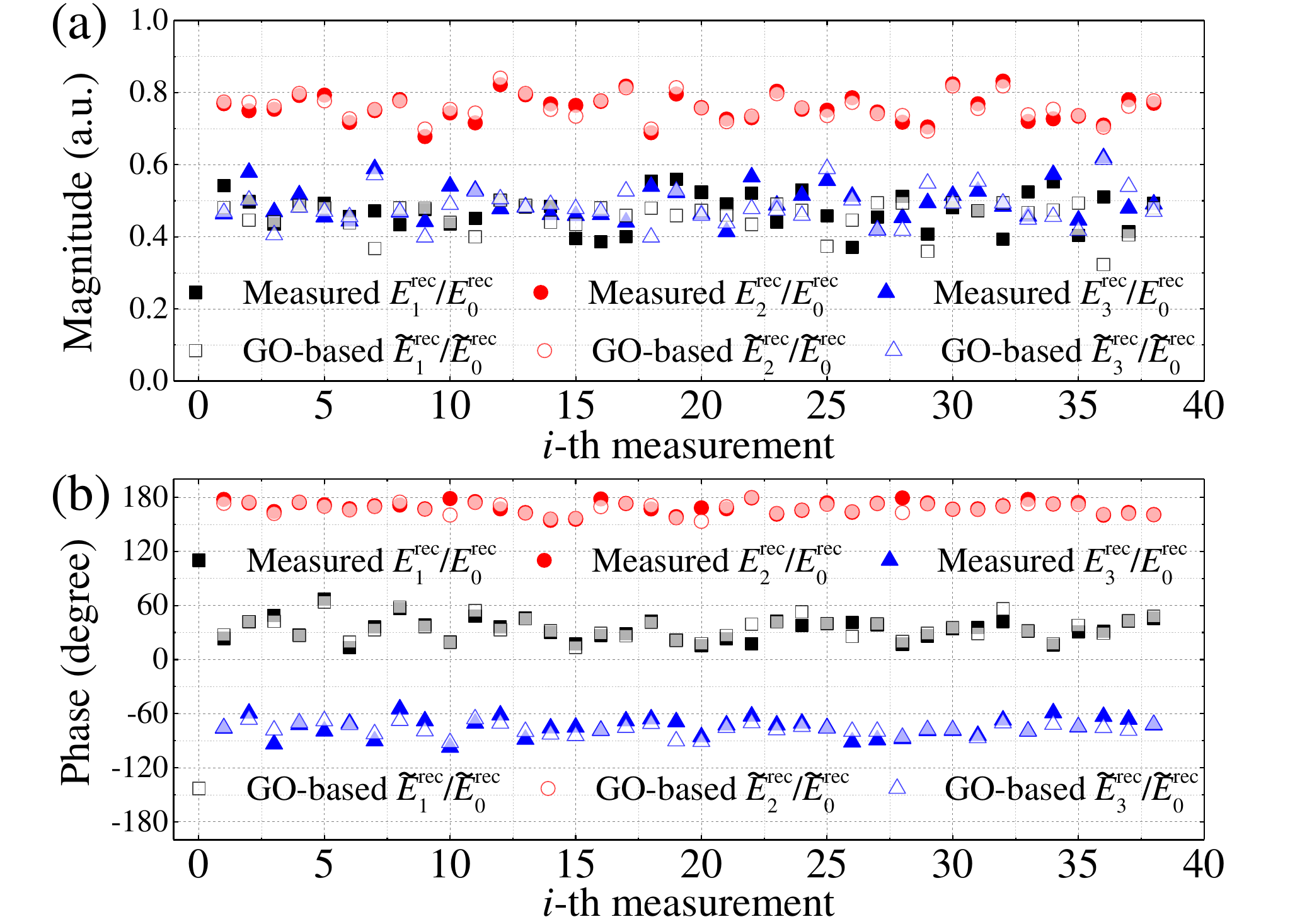}
\caption{The best matched magnitude (a) and phase (b) responses between the measured electric fields $E_{n}^{\mathrm{rec}}$ and GO predicted $\tilde E_{n}^{\mathrm{rec}}$, where the three considered focusing points are defined in Eq. (\ref{eq_position_focusingN}) with $N = 3$.}
\label{experiment_N123}
\end{figure}

To estimate the complex relative permittivity $\tilde \varepsilon_r = \tilde \varepsilon_r' - j\tilde \varepsilon_r''$ and a more precise thickness $\tilde T$ of the PA66 slab, $\varepsilon_r'$, $\varepsilon_r''$, and $T$ are swept to compute the GO predicted received amplitude $\tilde E_n^{{\rm{rec}}}(\varepsilon_r', \varepsilon_r'', T)$, which are then compared to the experimental measurements $E_n^{{\rm{rec}}}$ to find the best matched magnitude and phase responses using Eq.(\ref{eq_estimation_eps}). The measurement is performed $38$ times and 3 focusing points are considered. Figure \ref{experiment_N123} shows the best matched magnitude (a) and phase (b) responses between the measured electric fields $E_{n}^{\mathrm{rec}}$ and GO predicted $\tilde E_{n}^{\mathrm{rec}}({{\tilde \varepsilon _r'},\tilde \varepsilon _r'',\tilde T})$, corresponding to the minimum error $\mathrm{min}\left\{ f({{\varepsilon _r'},\varepsilon _r'',T}) \right\}$. As we can see, the received field magnitude $E_2^{{\rm{rec}}}$, corresponding to the focusing point at the range $z = 780$ mm, has the maximum magnitude compared to the measured magnitudes of the other focusing points along the $z$-axis. This is accordance with the experimentally reconstructed target profile, shown in Fig. \ref{experiment_profile2}, where the imaged object front surface is located at the range $z = 780$ mm.

\begin{figure}[t]
\centering
\includegraphics[width=\linewidth]{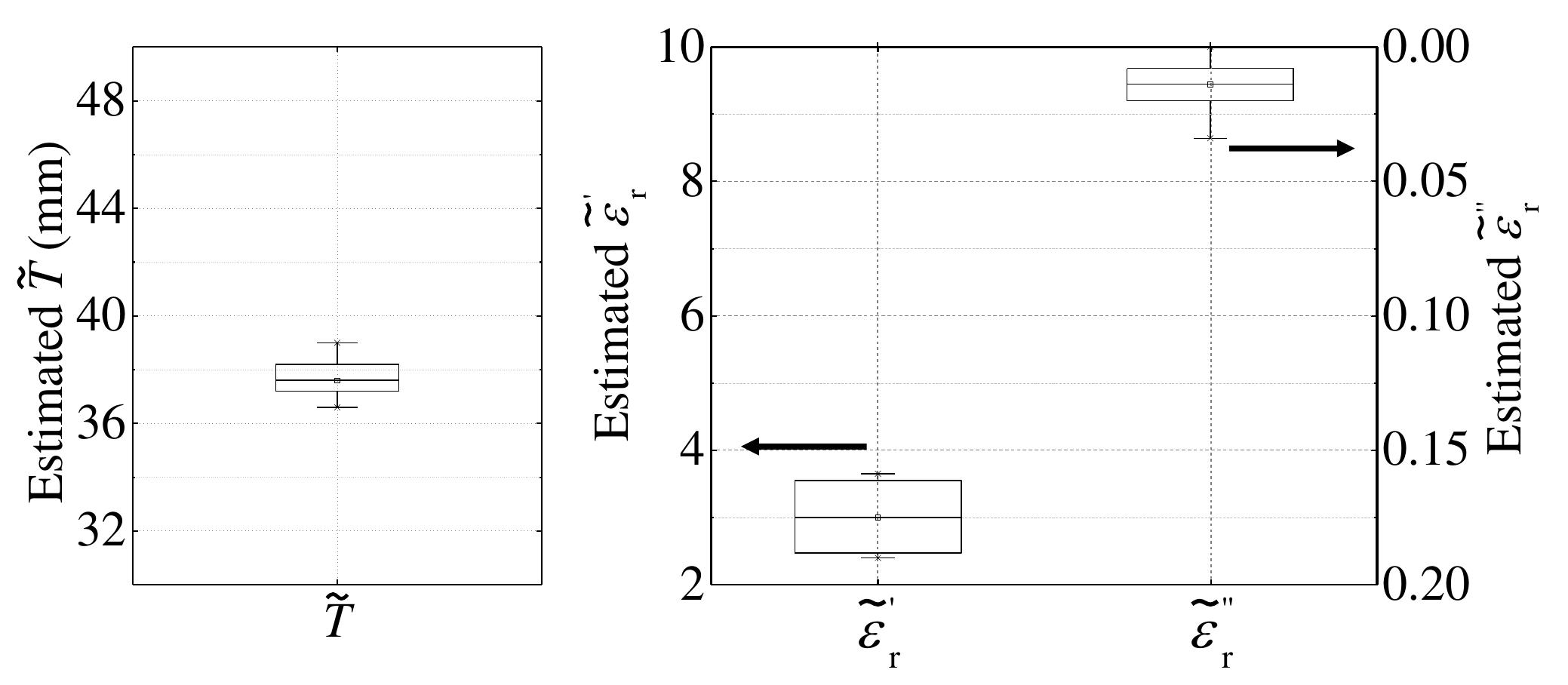}
\caption{Experimentally estimated dielectric constant $\tilde \varepsilon_r'$, loss factor $\tilde \varepsilon_r''$, and thickness $\tilde T$ of the PA66 dielectric slab.}
\label{experiment_estimation}
\end{figure}

The estimated results on $\tilde \varepsilon_r'$, $\tilde \varepsilon_r''$, and $\tilde T$ of the PA66 slab are given in Fig. \ref{experiment_estimation}. As shown in Table \ref{tabel_estimation}, the mean values for $\tilde \varepsilon_r'$, $\tilde \varepsilon_r''$, and $\tilde T$ are $3.012$, $0.014$, and $37.6$ mm, respectively; and the standard deviations for $\tilde \varepsilon_r'$, $\tilde \varepsilon_r''$, and $\tilde T$ are $0.425$, $0.009$, and $0.593$ mm, respectively. The estimation error can be attributed to the noisy experimental measurement as well as the approximated model of the four-layer transmission line.
\begin{table}[!t]
\renewcommand{\arraystretch}{2.0}
\caption{Estimated results on the thickness $\tilde T$ and complex relative permittivity $\tilde \varepsilon_r = \tilde \varepsilon_r' - j\tilde \varepsilon_r''$ for the PA66 slab}.
\centering
\begin{tabular}{c c c c}
\hline
\hline
Values & $\tilde T$ & $\tilde \varepsilon_r'$ & $\tilde \varepsilon_r''$ \\
\hline
Mean & $37.6$ mm & $3.012$ & $0.014$\\
\hline
Standard deviation & $0.593$ mm & $0.425$ & $0.009$\\
\hline
\hline
\end{tabular}
\label{tabel_estimation}
\end{table}

\section{Conclusion} \label{sec_conclusion}
A physical and geometrical optics imaging algorithm is derived for profile reconstruction and material identification in multiple reconfigurable reflectarrays based people-screening systems. Both simulations and experimental validations are carried out to examine the feasibility. Preliminary results show that the imaging system is able to not only reconstruct the target profile, but also characterize the complex relative permittivity of the dielectric object.

When accessible to a well defined database including the knowledge of the permittivities of typical threat materials, the proposed imaging scheme is capable of suggesting threat identities based on the estimated permittivities. This imaging scheme can have a variety of applications in security screening checkpoints at train stations, airports, concerts, sporting events, government buildings, and many other public and private facilities to predict potential threats.

\section*{Acknowledgment}
This work is funded by the U.S. Department of Homeland Security, Award No. 2013-ST-061-ED0001. The authors would like to thank Smiths Detection for collecting the data with their Eqo system.


\end{document}